\documentclass[english]{article}
\usepackage[T1]{fontenc}
\usepackage[latin9]{inputenc}
\usepackage{geometry}
\usepackage{babel}
\usepackage{float}
\usepackage{amsmath}
\usepackage{amsthm}
\usepackage{amssymb}
\usepackage{graphicx}
\usepackage{setspace}
\usepackage[unicode=true]
 {hyperref}

\makeatletter

\providecommand{\tabularnewline}{\\}
\floatstyle{ruled}
\newfloat{algorithm}{tbp}{loa}
\providecommand{\algorithmname}{Algorithm}
\floatname{algorithm}{\protect\algorithmname}

\theoremstyle{plain}
\newtheorem{thm}{\protect\theoremname}
\theoremstyle{definition}
\newtheorem{defn}[thm]{\protect\definitionname}
\theoremstyle{plain}
\newtheorem{prop}[thm]{\protect\propositionname}
\ifx\proof\undefined
\newenvironment{proof}[1][\protect\proofname]{\par
	\normalfont\topsep6\p@\@plus6\p@\relax
	\trivlist
	\itemindent\parindent
	\item[\hskip\labelsep\scshape #1]\ignorespaces
}{%
	\endtrivlist\@endpefalse
}
\providecommand{\proofname}{Proof}
\fi
\newenvironment{lyxlist}[1]
	{\begin{list}{}
		{\settowidth{\labelwidth}{#1}
		 \setlength{\leftmargin}{\labelwidth}
		 \addtolength{\leftmargin}{\labelsep}
		 }}
	{\end{list}}
\theoremstyle{plain}
\newtheorem{lem}[thm]{\protect\lemmaname}
\theoremstyle{plain}
\newtheorem{cor}[thm]{\protect\corollaryname}

\usepackage{colortbl}

\@ifundefined{showcaptionsetup}{}{%
 \PassOptionsToPackage{caption=false}{subfig}}
\usepackage{subfig}
\makeatother

\providecommand{\corollaryname}{Corollary}
\providecommand{\definitionname}{Definition}
\providecommand{\lemmaname}{Lemma}
\providecommand{\propositionname}{Proposition}
\providecommand{\theoremname}{Theorem}

\begin{document}
\title{K-bMOM: a robust Lloyd-type clustering algorithm based on bootstrap
Median-of-Means}
\author{Camille Brunet-Saumard\thanks{twice.ai, France.}, Edouard Genetay\thanks{Univ Rennes, Ensai, CNRS, CREST {[}(Center for Research in Economics
and Statistics){]} - UMR 9194, F-35000 Rennes, France and LumenAI,
France.}, Adrien Saumard\thanks{Univ Rennes, Ensai, CNRS, CREST {[}(Center for Research in Economics
and Statistics){]} - UMR 9194, F-35000 Rennes, France.}}
\maketitle
\begin{abstract}
We propose a new clustering algorithm that is robust to the presence
of outliers in the dataset. We perform Lloyd-type iterations with
robust estimates of the centroids. More precisely, we build on the
idea of median-of-means statistics to estimate the centroids, but
allow for replacement while constructing the blocks. We call this
methodology the bootstrap median-of-means (bMOM) and prove that if
enough blocks are generated through the bootstrap sampling, then it
has a better breakdown point for mean estimation than the classical
median-of-means (MOM), where the blocks form a partition of the dataset.
From a clustering perspective, bMOM enables to take many blocks of
a desired size, thus avoiding possible disappearance of clusters in
some blocks, a pitfall that can occur for the partition-based generation
of blocks of the classical median-of-means. Experiments on simulated
datasets show that the proposed approach, called K-bMOM, performs
better than existing robust K-means based methods. Guidelines are
provided for tuning the hyper-parameters K-bMOM in practice. It is
also recommended to the practitionner to use such a robust approach
to initialize their clustering algorithm. Finally, considering a simplified
and theoretical version of our estimator, we prove its robustness
to adversarial contamination by deriving robust rates of convergence
for the K-means distorsion. To our knowledge, it is the first result
of this kind for the K-means distorsion.
\end{abstract}

\section{Introduction}

\sloppypar Data scientists have nowadays to deal with massive and
complex datasets, that are often corrupted by outliers. Classical
data mining procedures such as K-means or more general EM algorithms
for instance are however sensitive to the presence of outliers, which
can induce a time consuming pre-processing of the data.

In this context, robust versions of data mining procedures are particularly
relevant and we investigate a way to produce a Lloyd-type algorithm
for hard clustering that is robust to the presence of ouliers. We
propose more precisely to use a variant of median-of-means (MOM) statistics,
that we call ``bootstrap median-of-means'' (bMOM). MOM principle
has been the object of recent intensive research in mean estimation,
regression, high-dimensional framework and also supervised classification
and machine learning (\cite{lerasle2011robust,MR3576558,lecue2017learning,lecue2017robust,MR3909950,MR4017683,MR4055993,minsker2018uniform}).
It is worth noting that other approaches to robustness for K-means
exist in the literature, such as for instance K-median or trimmed
K-means (see for instance the survey \cite{MR2720402} and references
therein ; see also \cite{brecheteau2018robust}).

Given a dataset, the boostrap median-of-means consists in first generating
a (large) bootstrap sample and then perform a classical median-of-means
on this bootstrap sample. We prove in Section \ref{sec:Robust-mean-estimation}
that if enough blocks are generated from the bootstrap sampling, then
for a fixed block size, bMOM has a higher breakdown point than MOM. 

We propose a robust-to-outliers version of K-means, that we call K-bMOM,
and that performs Lloyd-type iterations through the use of bMOM estimates
of the K-means distorsion, as further explained in Section \ref{sec:kB-MOM-algorithm}. 

We provide in Section \ref{sec:Theoretical-analysis} some deviation
bounds for the performance in terms of K-means distorsion of an idealized
version of the estimator produced by our algorithm. We consider indeed
a minimizer of the median-of-means of the K-means distorsion loss
along possible codebooks. We prove that such an estimator is robust
to adversarial contamination of the dataset if the number of outliers
is sufficiently small compared to the number of blocks in the MOM
statistics. 

In Section \ref{sec:Simulations-and-practical}, a bMOM based procedure
is considered to initialize clustering algorithms and is compared
to existing initialisation on simulated datasets. Practical considerations
to choose the number and size of blocks are discussed and guidelines
are provided. Finally, the K-bMOM algorithm is compared to existing
robust K-means based clustering approach on simulated datasets with
the presence of outliers.

Finally, we note that our framework is close to the recent work \cite{Klochkov2020robust}
that investigates the use of median-of-means statistics to produce
a robust K-means type clustering. However, the latter work is theoretical
only and the authors study probabilistic performance bounds for the
minimizer of the median-of-means of the K-means distorsion loss under
a finite second moment assumption. In particular the authors do not
discuss the use of median-of-means through Lloyd-type iterations nor
a practical way to compute the estimator. Neither do they discuss
the possibility of generating blocks with replacements in the dataset.

\section{Robust mean estimation by the bootstrap median-of-means\label{sec:Robust-mean-estimation}}

\subsection{Median-of-Means and bootstrap Median-of-Means}

The median-of-means (MOM) estimator of the mean in dimension one consists
in taking a median of some arithmetic means computed on a collection
- say of size $B$ - of disjoint blocks $\left(x_{i}\right)_{i\in b_{k}}$,
where $\left\{ b_{k}:k\in\left\{ 1,...,B\right\} \right\} $ form
a partition of the set of indices $\left\{ \text{1,...,n}\right\} $
of a real valued sample $x_{1}^{n}=\left(x_{1},...,x_{n}\right)$.
The length of the blocks are generally taken to be equal, eventually
up to one data. We can thus write, by denoting $b_{1}^{B}$ the collection
of blocks,
\[
{\rm MOM}(x_{1}^{n},b_{1}^{B})={\rm med}\left\{ \sum_{j\in b_{k}}x_{j}:k\in\left\{ 1,...,B\right\} \right\} .
\]
where ${\rm med}$ is a median, that is $\#\left\{ k\in\left\{ 1,...,B\right\} ;a_{k}\leq{\rm med}\left\{ a_{i}\right\} \right\} \geq B/2$
and $\#\left\{ k\in\left\{ 1,...,B\right\} ;a_{k}\geq{\rm med}\left\{ a_{i}\right\} \right\} \geq B/2$. 

We may consider that the blocks are generated according to a random
drawing process, that proceeds whithout replacements (disjoint blocks)
and according to the uniform distribution on the remaining data at
each step. This formulation naturally leads to consider more general
random block generating processes.

For any positive integers $n_{B}$ and $B$, denote $m=Bn_{B}$ and
generate a bootstrap sample $y_{1}^{m}=(y_{1},...,y_{m})$ from the
dataset $x_{1}^{n}$. More precisely, each $y_{i}$ is taken uniformly
at random from the values $\left(x_{1},...,x_{n}\right)$ and independently
from the $\left(y_{j}\right)_{j\neq i}$. Then the boostrap median-of-means
(bMOM) of the dataset $x_{1}^{n}$ with parameters $n_{B}$ and $B$
is the (classical) MOM estimator on the boostrap sample $y_{1}^{m}$
with blocks $b_{j}=(n_{B}(j-1)+1,...,n_{B}j)$ for $j\in\{1,...,B\}$,
\[
{\rm bMOM}(x_{1}^{n},n_{B},B)={\rm MOM}(y_{1}^{m},b_{1}^{B}).
\]

It is worth noting that ${\rm bMOM}$ is a randomized estimator. Also,
for any fixed sample size $n$, we can choose any block size $n_{B}$
and number of blocks $B$ to define a ${\rm bMOM}$ estimator, on
contrary to the classical MOM, where the product of the block size
with the number of blocks should be equal to the sample size. This
will turn out to be precious in the clustering context, where we do
not want too small sample block sizes in order to avoid disappearance
of some clusters in the blocks.

We prove below that taking enough blocks in the definition of bMOM
enables to perform a more robust estimation than with MOM and same
block size, in the sense that the breakdown point of the bMOM is higher.
This also provides an interest to bMOM compared to MOM for mean estimation
in general. We leave as an interesting open problem the question of
sub-gaussian deviation bounds, in the flavor of \cite{MR3576558},
for mean estimation using bMOM.

\subsection{Breakdown points\label{subsec:Breakdown-points}}

The breakdown point is a classical concept in the robust statistics
literature (\cite{MR2488795,MR3839299}), that gives the maximal proportion
of outliers that is allowed so that the deviations of the estimator
stay bounded compared to the no-corruption setting.

Assume that we are given a sample $x_{1}^{n}=$$\left(x_{1},...,x_{n}\right)$
of real valued random variables.
\begin{defn}[Deterministic Breakdown point]
\label{def:The-(deterministic)-breakdown}The (deterministic) breakdown
point $\delta_{n}\left(T_{n},x_{1}^{n}\right)$ of an estimator $T_{n}$
given the sample $x_{1}^{n}$ is the maximal proportion of outliers
that leave the value of the estimator bounded.
\[
\delta_{n}\left(T_{n},x_{1}^{n}\right)=\frac{1}{n}\max\left\{ m;\max_{i_{1},...,i_{m}}\sup_{y_{1},...,y_{m}}\left|T_{n}\left(z_{1},...,z_{n}\right)\right|<+\infty\right\} \,,
\]
where the sample $\left(z_{1},...,z_{n}\right)$ is obtained by replacing
the $m$ data points $x_{i_{1}},...,x_{i_{m}}$ of the sample $x_{1}^{n}$
by arbitrary values $y_{1},...,y_{m}$.
\end{defn}
One can notice that Definition \ref{def:The-(deterministic)-breakdown}
corresponds to a worst case analysis, the outliers potentially appearing
at the worst places for the estimator $T_{n}$. If the estimator $T_{n}$
is randomized - we rather denote it $T_{n}^{\omega}$ in this case
-, then its breakdown point is a random variable.

For a median ${\rm med}(x_{1}^{n})$, it holds $\delta_{n}\left({\rm med}(x_{1}^{n}),x_{1}^{n}\right)$=$\left\lfloor n/2\right\rfloor /n$
and for the empirical mean $\bar{x}_{n}=1/n\sum_{i=1}^{n}x_{i}$,
$\delta_{n}\left(\bar{x}_{n},x_{1}^{n}\right)=1/n$. For the median-of-means
estimator, 
\[
\delta_{n}\left({\rm MOM}(x_{1}^{n},b_{1}^{B}),x_{1}^{n}\right)=\left\lfloor B/2\right\rfloor /n\;a.s.,
\]
since it suffices to have one outlier in a majority of blocks to make
MOM diverge. 

Note that \cite[Section 4.2]{MR3576558} proposes to automatically
select the number of blocks of the MOM estimator by a Lepskii-type
procedure that consists in choosing the smallest number of blocks
such that the intersection of some confidence intervals constructed
for MOM with greater numbers of blocks is empty. The resulting estimator
will inherit from the value of the breakdown point corresponding to
the highest number of blocks in the considered collection. If the
highest number of blocks is $n$, the sample size, thus corresponding
to a median, then the method of intersection of confidence intervals
gives an optimal value of breakdown point, corresponding $\left\lfloor n/2\right\rfloor /n$. 

However, computing such selection procedure is time consuming and
as we want to make an iterative use of (bootstrap) MOM estimates,
this method seems to be out of the scope for us. Instead, we show
below that the use of replacements while constructing the blocks already
gives an improvement of the breakdown point if enough blocks are considered,
compared to the use of disjoint blocks when applied to MOM statistics.
\begin{prop}
\label{prop:deterministic_bdp}We have
\[
\delta_{n}\left({\rm bMOM}(x_{1}^{n},n_{B},B),x_{1}^{n}\right)\leq\delta_{n}\left({\rm MOM}(x_{1}^{n},b_{1}^{B}),x_{1}^{n}\right)\;a.s.
\]
and, for a fixed parameter block size $n_{B}$,
\[
\lim_{B\rightarrow+\infty}\delta_{n}\left({\rm bMOM}(x_{1}^{n},n_{B},B),x_{1}^{n}\right)=1-\frac{1}{2^{1/n_{B}}}>\frac{1}{2n_{B}}\,\;a.s.
\]
Note that $1-\frac{1}{2^{1/n_{B}}}\sim_{n_{B}\rightarrow+\infty}\frac{\log2}{n_{B}}\simeq\frac{0.69}{n_{B}}$.
\end{prop}
On the one hand, the first display in Proposition \ref{prop:deterministic_bdp}
states that when the number of blocks in bMOM is equal to the number
of blocks in MOM, bMOM has a breakdown point that is smaller than
or equal to the breakdown point of MOM (this is due to the possible
repetitions of outliers along the blocks for bMOM). On the other hand,
the second display in Proposition \ref{prop:deterministic_bdp} states
that for a fixed block size, when the number of blocks in bMOM tends
to infinity, its breakdown point tends to a value that is strictly
greater than the breakdown point of MOM with the same block size. 
\begin{proof}
For the second display. Assume that the sample is corrupted by $m$
outliers. Denote $S_{i}$ the indicator that the block $B_{i}$ is
not corrupted. Then $S_{i}$ is a Bernoulli random variable of mean
$\left(1-m/n\right)^{n_{B}}.$ Then $\sup_{y_{1},...,y_{m}}\left|{\rm bMOM}(x_{1}^{n},n_{B},B)\right|$
is finite if the proportion of corrupted blocks smaller than $1/2$.
This corresponds to the condition $\sum_{i=1}^{B}S_{i}/B>1/2$. By
the strong law of large numbers, the latter is almost surely realized
asymptotically if $\left(1-m/n\right)^{n_{B}}>1/2$, hence the result.
\end{proof}
Considering that the contaminated sample is given (fixed), it is interesting
to evaluate the probability that a randomized estimator does not diverge
when the outliers go to infinity. It can indeed happen that the indices
of the outliers are not the worst with respect to the block drawing
process. This leads to the following definition.
\begin{defn}[Probabilistic Breakdown point]
 The probabilistic breakdown point of a randomized estimator $T_{n}^{\omega}$
given the sample $x_{1}^{n}$ is
\end{defn}
\[
p_{n}\left(T_{n}^{\omega},x_{1}^{n},\left(i_{1},...,i_{m}\right)\right)=\mathbb{P}\left(\left\{ \omega:\sup_{y_{1},...,y_{m}}\left|T_{n}^{\omega}\left(z_{1},...,z_{n}\right)\right|<+\infty\right\} \right)
\]
where the sample $\left(z_{1},...,z_{n}\right)$ is obtained by replacing
the $m$ data points $x_{i_{1}},...,x_{i_{m}}$, for some fixed indices
$\left(i_{1},...,i_{m}\right)$, by the arbitrary values $y_{1},...,y_{m}$. 

\begin{figure}
\begin{centering}
\includegraphics[scale=0.5]{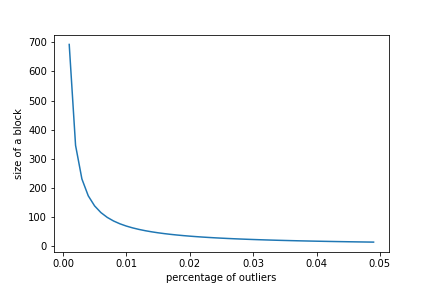}
\par\end{centering}
\caption{Maximum admissible block size $n_{B}$ according to the proportion
of outliers $p=\frac{m}{n}$ \label{fig:Size of block according to proportion of outliers}}

\end{figure}

\begin{figure}
\begin{centering}
\includegraphics[scale=0.5]{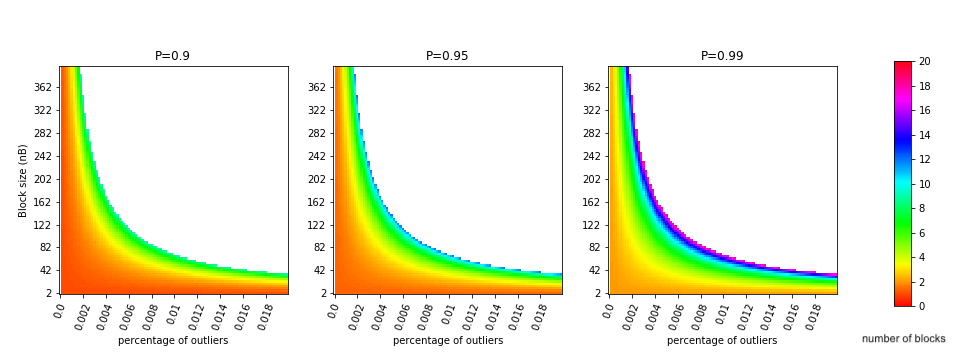}
\par\end{centering}
\caption{Evolution of the lower bound on the number of blocks (colorbar) according
to the proportion of outliers and the size of the blocks for different
levels of confidence.\label{fig:Lower bound of the nb of blocks according to nB and p}}

\end{figure}

As $p_{n}\left({\rm bMOM}(x_{1}^{n},n_{B},B),x_{1}^{n},\left(i_{1},...,i_{m}\right)\right)$
only depends on $n$ and $m$, but not on the values of $\left(i_{1},...,i_{m}\right)$
or $x_{1}^{n}$, we will rather denote it $p_{n}\left({\rm bMOM}(x_{1}^{n},n_{B},B),m\right)$.
We have the following bound.
\begin{prop}
It holds
\[
p_{n}\left({\rm bMOM}(x_{1}^{n},n_{B},B),m\right)\geq1-\exp\left(-2B\left(\left(1-m/n\right)^{n_{B}}-1/2\right)^{2}\right)\,.
\]
\end{prop}
If the number of outliers $m$ and the sample size $n$ are fixed
then the block length $n_{B}$ should be such that $\left(1-m/n\right)^{n_{B}}>1/2$,
that is $n_{B}<\log(2)/\log\left(\left(1-m/n\right)^{-1}\right)$.
Hence, in case of a large proportion of outliers $m/n$, the block
length should not be taken too large (see Figure~\ref{fig:Size of block according to proportion of outliers}
to visualize the previous condition on the block size according to
the proportion of outliers). Furthermore, by denoting $D=$$\left(1-m/n\right)^{n_{B}}-1/2(>0)$,
we have that $p_{n}\left({\rm bMOM}(x_{1}^{n},n_{B},B),m\right)\geq1-R$
is equivalent to $B>\log\left(1/R\right)/\left(2D^{2}\right)$. We
illustrate the behavior of the latter lower bound on the block size
in Figure~\ref{fig:Lower bound of the nb of blocks according to nB and p}.
This implies in particular that if the block size $n_{B}$ is rightly
chosen (not too large according to the proportion of outliers), then
the probability that the bootstrap median-of-means remains stable
under the adversarial contamination tends to one when the number of
blocks $B$ tends to infinity. 
\begin{proof}
As in the proof of Proposition \ref{prop:deterministic_bdp}, denote
$S_{i}$ the indicator that the block $B_{i}$ is not corrupted. We
have, by Hoeffding's inequality (\cite[Theorem 2.27]{MR3363542}),
\begin{align*}
\mathbb{P}\left(\left\{ \omega:\sup_{y_{1},...,y_{m}}\left|{\rm bMOM}(x_{1}^{n},n_{B},B)\right|=+\infty\right\} \right) & =\mathbb{P}\left(\sum_{i=1}^{B}\left(1-S_{i}\right)>B/2\right)\\
 & \leq\exp\left(-2B\left(\left(1-m/n\right)^{n_{B}}-1/2\right)^{2}\right).\\
\end{align*}
\end{proof}

\section{$K$-bMOM algorithm\label{sec:kB-MOM-algorithm}}

We propose in this section an estimation procedure based on bMOM statistics
for clustering unlabeled data. Moreover, since the resulting partition
of most of clustering approaches depends on the starting centers,
we propose also a bMOM-based initialization procedure.

Let us introduce the following notations. Let $x_{1},\dots,x_{n}\in\mathbb{R}^{p}$
denote a dataset of $n$ observations that we want to cluster into
$K$ homogeneous groups. Then $b\in\{1,\dots,B\}$ stands for the
index of a block $b$ and $B\in\mathbb{N^{*}}$ the number of blocks,
containing at least $n_{B}>K$ datapoints. We define the empirical
risk of the block $b$ as:
\[
R_{b}(\textbf{c})=\sum_{k=1}^{K}\sum_{i\in\mathcal{C}_{k}^{(b)}}\left\Vert x_{i}^{(b)}-c_{k}^{(b)}\right\Vert ^{2}
\]
where $x_{i}^{(b)}$ stands for the $i$th datapoint contained in
the block $b$, $\mathcal{C}_{k}^{(b)}$ stands for the set of datapoints
belonging to cluster $k$ in the block $b$ and $\left\Vert .\right\Vert $
is the Euclidean norm. Furthermore, $c_{k}^{(b)}$ stands for the
mean vector of the cluster $k$ in the block $b$ and we denote by
$v_{k}^{(b)}$ its within variance. Finally, we denote by $\mathcal{P}(\mathbf{c})$
the Vorono� partition obtained from the set of centroids~$\mathbf{c}$.

\subsubsection*{A robust initialisation}

It is well-known that since the clustering problem is non convex,
the initialisation step is a keystone for the resulting partition.
We propose therefore a robust variant of traditional initialisation
strategies by applying the MOM principle. To do so, the idea is to
build uniformly and with replacement $B$ blocks of $n_{B}$ datapoints
where the number of points is strictly greater than the number of
groups. In each block a traditional K-means++ initialisation~\cite{Vassilvitskii07}
is operated. Such an approach proceeds iteratively: it starts with
a centroid picked at random among the datapoints. Then, iteratively
and until the number of groups $K$ is reached, a new centroid is
chosen from the datapoints with a probability which increases exponentially
with the distance $D^{2}(x,c)$ to the closest centers already chosen.
In each block, the empirical risk is therefore computed and the centers
linked to the median empirical risk, called the median block, is selected
as the initial centers.

We define the following algorithm:

\begin{algorithm}[H]
\textbf{Input:} the dataset $\left\{ x_{1},\ldots,x_{n}\right\} $,
$B$ the number of blocks and $n_{B}>K$ size of blocks
\begin{enumerate}
\item Iterate from $1$ until $B$ blocks:
\begin{enumerate}
\item Select at random, uniformly and with replacement $n_{B}$ datapoints
\item Proceed a kmeans++ initialisation
\item Compute the empirical risk $\hat{R}_{b}\mathrm{(\textbf{c})}$ of
the block $b$
\end{enumerate}
\item Select the centers from the block having the median empirical risk
and get: $\left(\widehat{c}_{bmed}^{(1)},\ldots,\widehat{c}_{bmed}^{(K)}\right)$.
\end{enumerate}
\textbf{Output:} $\left(\widehat{c}_{bmed}^{(1)},\ldots,\widehat{c}_{bmed}^{(K)}\right)$

\caption{Initialisation strategy\label{alg:Initialisation-strategy}}
\end{algorithm}

\subsubsection*{The K-bMOM algorithm}

Due to the nature of the bMOM statistics and the clustering goal,
the algorithm that we propose alternates three main steps. At iteration
$t$, and given the centers fitted in the median block of the previous
iteration, $B$ blocks of $n_{B}$ data are built by uniform sampling
with replacement. Then, a partition per block is computed by assigning
each data point to its closest centroids fitted on the median block
at iteration ($t-1$). The centroids of each block are updated according
to their block partition and the empirical risk $\hat{R}_{b}\mathrm{(\textbf{c})}$
is returned. The block with the median empirical risk is selected
and the fitted centers of this median block become the current ones.
This is done until the empirical risk of the median block $\hat{R}_{bmed}(\textbf{c})$
remains stable. The final partition on all the dataset is obtained
by assigning each data point to its nearest closest centroid $\left(\widehat{c}_{1}^{(bmed)},\dots,\hat{c}_{K}^{(bmed)}\right)$
of the current median block. 

A pseudo algorithm of this procedure is detailed in Algorithm~\ref{alg:Iteration-phase-structure-1}.

\bigskip

\begin{algorithm}[h]
\caption{Iteration phase structure \label{alg:Iteration-phase-structure-1}}

\textbf{Input:} $\left\{ x_{1},\ldots,x_{n}\right\} $, $B$ the number
of blocks and $n_{B}$ size of blocks ($n_{B}$>$K$)

\textbf{Initialisation step:} Algorithm~\ref{alg:Initialisation-strategy}.

\textbf{Set:} $q=0$ and $crit>>\varepsilon$.

\medskip

\textbf{Main Loop:} while $crit>\varepsilon$ or $q<\bar{q}_{max}$:
\begin{enumerate}
\item Create $B$ blocks of the data of size $n_{B}$ randomly and uniformly
with replacement
\item In each block $b$:
\begin{itemize}
\item Assign each datapoint to its closest centroid.
\item If $n_{k}^{(b)}>1,$ $\forall k\in\{1,\dots,K\}$:
\begin{itemize}
\item for $k\in\{1,\dots,K\}$: $c_{k}^{(b)}\leftarrow1/n_{k}^{(b)}\sum_{i\in\mathcal{C}_{k}}x_{i}^{(b)}$
\item $\hat{R}_{b}\mathrm{(\textbf{c})}\leftarrow\sum_{k=1}^{K}\sum_{i\in\mathcal{C}_{k}^{(b)}}\left\Vert x_{i}^{(b)}-\hat{c}_{k}^{(b)}\right\Vert ^{2}$
\end{itemize}
\end{itemize}
\item Get the median empirical risk $\hat{R}_{bmed}\mathrm{(\textbf{c})}$
and the associated quantities of the median block\,: $b_{med}$,
$\left(\widehat{c}_{1}^{(bmed)},\ldots,\widehat{c}_{K}^{(bmed)}\right)$.
\item $q\leftarrow q+1$
\end{enumerate}
\medskip
\begin{description}
\item [{Output:}] $\left(\bar{c}_{bmed}^{(1)},\ldots,\bar{c}_{bmed}^{(K)}\right)$
and $\mathcal{P}(\hat{\mathbf{c}}_{med})$
\end{description}
\end{algorithm}

\subsubsection*{Stopping criterion}

In practice, the algorithm is run a given number of maximum iterations
($\bar{q}_{max}$ = 50 by default). In order to obtain a more precise
estimation of centroids at the end of the maximum number of iterations,
instead of retrieving the centroids of the median block computed in
the last iteration, centroids of the last 10 iterations are agregated
$\left(\bar{c}_{bmed}^{(1)},\ldots,\bar{c}_{bmed}^{(K)}\right)$.

\subsubsection*{Model selection}

In model-based clustering, it is frequent to consider several models
in order to find the most appropriate one for the considered data.
In particular, for most of clustering algorithms, the model is specified
by its number of clusters $K$. There are lots of ad-hoc approaches
in the literature to select the number of components $K$ and we can
therefore think of the Gap statistics from~\cite{tibshirani2001estimating},
the Silhouette criterion and so one. However, since the K-means algorithm
can be seen as a hard version of an EM-like algorithm which tries
to estimate a mixture of $K$ Gaussians with isotropic covariance
matrices, we can therefore apply classical tools for model selection
including BIC, ICL criteria and the heuristic slope~\cite{BauMauMich:12}
for example. We can therefore use such criteria on the proposed robust
version of the K-means by processing the K-bMOM on several values
of $K$, computing the chosen criterion for each model and select
the model defined by its number of components which either maximizes
the BIC or ICL criteria or follow the principle of the slope heuristic.

\section{Theoretical analysis\label{sec:Theoretical-analysis}}

In this section, we give probabilistic performance bounds for a theoretical
and simplified version of the estimator produced by our algorithm
presented in Section \ref{sec:kB-MOM-algorithm} above.

We need first to describe our setting. We study the \textit{robustness
against adversarial contamination}. Since we are in a probabilistic
framework, we denote the sample $\left(X_{1},...,X_{n}\right)$, rather
than $\left(x_{1},...,x_{n}\right)$ in the previous sections. We
assume that the dataset is made of two disjoint components: the set
of inliers $(X_{i})_{i\in\mathcal{I}}$, corresponding to data that
bring information and are not corrupted, and the set of outliers $(X_{j})_{j\in\mathcal{O}}$,
that may be completely misleading for the clustering task. The random
variables $X_{i}$, $i=1,...,n$, take values in a separable Hilbert
space $\left(\mathcal{X},\left\Vert \cdot\right\Vert \right)$ and
the inliers $(X_{i})_{i\in\mathcal{I}}$ are independent and identically
distributed random variables. No assumption is made on the behavior
of the outliers $(X_{j})_{j\in\mathcal{O}}$.

We also set a generic random variable $X$, independent from the sample
and of the same distribution $P$ as $X_{i}$, for any index $i\in\mathcal{I}$. 

For any codebook $\mathbf{c}=\{c_{1},...,c_{k}\}$, we denote by $\ell_{\mathbf{c}}$
a loss function on $\mathcal{X}$ such that $\ell_{\mathbf{c}}\left(x\right)=\min_{j=1,...,k}\left\{ -2\left\langle x,c_{j}\right\rangle +\|c_{j}\|^{2}\right\} $,
where $\left\langle \cdot,\cdot\right\rangle $ is the scalar product
associated to the Hilbertian norm $\left\Vert \cdot\right\Vert $
on $\mathcal{X}$. Notice that $\|x-c_{j}\|^{2}=\|x\|^{2}-2\left\langle x,c_{j}\right\rangle +\|c_{j}\|^{2}$.
The loss $\ell_{\mathbf{c}}$ is classically associated to the K-means
procedure (see for instance \cite{MR2444554}).

For any function $f$, denote $Pf:=\mathbb{E}\left[f\left(X\right)\right]$.
For the K-means problem to make sense, we assume that $P\|X\|^{2}<+\infty$.
Our goal is to find from the sample $(X_{1},...,X_{n})$ a collection
of centroids that is close to the following set of optimal codebooks,
\begin{align*}
C_{*} & =\arg\min_{\mathbf{c}\in\mathcal{X}^{k}}\left\{ P\ell_{\mathbf{c}}\right\} \\
 & =\arg\min_{\mathbf{c}=\{c_{1},...,c_{k}\}\in\mathcal{X}^{k}}\left\{ \mathbb{E}\left[\min_{j=1,...,k}\|X-c_{j}\|^{2}\right]\right\} .
\end{align*}
Also denote $\ell_{*}=\ell_{\mathbf{c}_{*}}$ for any $\mathbf{c}_{*}\in C_{*}$,
the optimal distorsion risk.

Furthermore, we assume that the magnitude of an optimal codebook is
known. This means that there exists a constant $M_{*}>0$ such that
there exists $\mathbf{c}_{*}=(c_{*,1},...,c_{*,k})\in C_{*}$ with
$\max_{i=1,...,k}\left\Vert c_{*,i}\right\Vert \leq M_{*}$ and that
we may restrict our search within codebooks $\mathbf{c}$ satisfying
$\max_{c\in\mathbf{c}}\left\Vert c\right\Vert \leq M_{*}$. 

Hence, we set
\begin{equation}
\hat{C}=\arg\min_{\mathbf{c}\in\mathcal{X}_{M_{*}}^{k}}\left\{ {\rm MOM}\left(\ell_{\mathbf{c}}\right)\right\} ,\label{eq:def_emp_min}
\end{equation}
the set of codebooks minimizing the median-of-means of the loss along
the data, where $\mathcal{X}_{M_{*}}=\left\{ x\in\mathcal{X};\left\Vert x\right\Vert \leq M_{*}\right\} $
is the ball of radius $M_{*}$ in $\mathcal{X}$ and we recall that 

\[
{\rm MOM}\left(\ell_{\mathbf{c}}\right)={\rm med}\left\{ \sum_{j\in b_{i}}\ell_{\mathbf{c}}(X_{j}):i\in\left\{ 1,...,B\right\} \right\} .
\]
We consider that our algorithm, presented in Section \ref{sec:kB-MOM-algorithm}
above, is an approximation of the minimization task defined in (\ref{eq:def_emp_min}).
Indeed, our algorithm iteratively computes codebooks in a Lloyd-type
fashion in each block of data and then chooses to keep at each step
the codebook that achieves the median of the K-means distorsion in
each block. 

Note also that we consider in (\ref{eq:def_emp_min}) the ``classical''
MOM, instead of the bootstrap MOM. But considering a bMOM with the
same block length and number of blocks as a MOM should give rather
similar performances. The point in using the MOM statistics is that
its mathematical analysis is simpler than for the bMOM, since the
blocks of the MOM are disjoint and are so independent. By consequence,
empirical process techniques will be available.

It is worth noting that the estimators given by (\ref{eq:def_emp_min})
have been recently studied in \cite[Section 2]{Klochkov2020robust},
where they are proved to achieve sub-Gaussian performance bounds under
only a two finite moments assumption for the random variable $X$.
Our result below complement the analysis carried in \cite{Klochkov2020robust}
by studying robustness against adversarial contamination rather than
robustness to heavy tailed data. In the framework of supervised learning,
\cite{lecue1808robust} also studied estimators of the form of (\ref{eq:def_emp_min})
- but with different losses -, both in the cases of data with finite
second moment and data contamination.

Let $O$ denote the set of indexes of blocks that contain at least
one outlier and $I$ denote the set of indexes of blocks that are
not corrupted, that is that do not contain any outlier. We thus have
$|O|\leq n_{o}$, where $n_{o}$ is the number of outliers , and $|I|\geq B-n_{o}$.

Denote also $R(\mathbf{c})=P\ell_{\mathbf{c}}$ the risk of a codebook
$\mathbf{c}$ and $R_{*}=P\ell_{*}$ the best possible risk. For any
$\hat{\mathbf{c}}_{n}\in\hat{C}$, we give probabilistic bounds on
$R\left(\hat{\mathbf{c}}_{n}\right)-R_{*}$, also known as the excess
K-means distorsion risk.
\begin{thm}
\label{thm:perf_bound}If there exists $M_{I}>0$ such that $\left\Vert X\right\Vert \leq M_{I}$
$a.s.$ and if the number of outliers $n_{o}$ satisfies $n_{o}\leq B/4$,
then there exists two numerical constants $l_{1},l_{2}>0$ such that
it holds, with probability greater than $1-2\exp\left(-l_{1}B\right)$,
\begin{equation}
R\left(\hat{\mathbf{c}}_{n}\right)-R_{*}\leq l_{2}\max\left\{ M\sqrt{\frac{B\mathbb{E}\left[\Vert X\Vert^{2}\right]}{n}},\frac{k\left[M\sqrt{\mathbb{E}\left[\Vert X\Vert^{2}\right]}+M^{2}/2\right]}{\sqrt{n}}\right\} ,\label{eq:bound_thm}
\end{equation}
where $M=\max\left\{ M_{*},M_{I}\right\} $. It can be seen from the
proof that $l_{1}=3/64$ and $l_{2}=512$ work.
\end{thm}
The proof of Theorem \ref{thm:perf_bound} can be found in Section
\ref{sec:Proof-of-Theorem}.

Note that in Theorem \ref{thm:perf_bound} we assumed that the inliers
are defined in a bounded domain of the Hilbert space $\mathcal{X}$
and robustness is considered through the fact that there may be outliers
in the dataset. If the number of outliers is small enough compared
to the number of blocks ($n_{o}\leq B/4$), the upper bound given
in (\ref{eq:bound_thm}) for the excess K-means distorsion risk is
composed of two terms. The second term in the maximum appearing at
the right-hand side of (\ref{eq:bound_thm}) correponds to the classical
convergence rate of the K-means for a sample that is bounded in a
separable Hilbert space that do not contain any outlier, see \cite{MR2444554}.
The first term in the maximum appearing at the right-hand side of
(\ref{eq:bound_thm}) reflects the price to pay for the presence of
outliers. In particular, it does not change the rate of convergence
of the no-contamination setting if $B$ is of the order of $k^{2}$.

\section{Simulations and practical considerations\label{sec:Simulations-and-practical}}

\subsection{Comparing initialisation strategies for the clustering task\label{subsec:Comparing-initialisation-strategies}}

It is well-known that the resulting partition of most clustering approaches
such as for example the K-means or the Gaussian Mixture models, heavily
depends on the starting centers. Therefore, a bad initialisation leads
to a poor partitioning of the data. This is particularly true in the
context of data with outliers where most of traditional and state-of-the-art
initialisation techniques behave poorly in such a context. We propose
in this section to apply the MOM principle to the most widely used
initialisation methods among which kmeans++ and kmedians++ . We evaluate
and compare them to their traditional use. 

These different strategies will be compared on simulated data in two
different contexts of outliers: punctual, spread out outliers and
a cluster of outliers. 

\paragraph{Simulation contexts:}

\begin{figure}[h]
\begin{centering}
\subfloat[Data generated without outliers]{\includegraphics[scale=0.29]{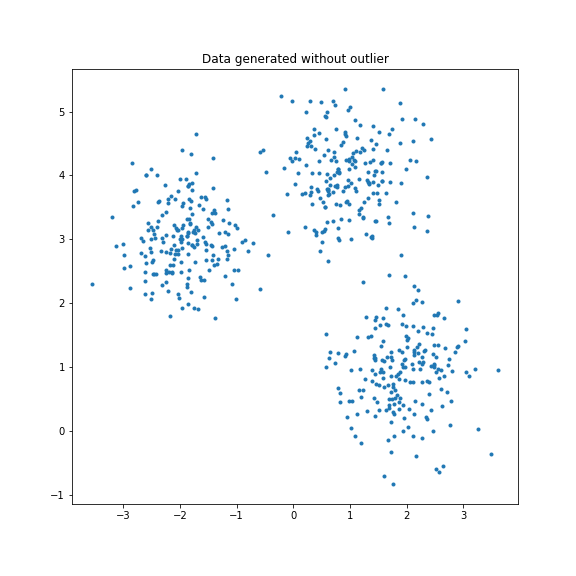}

}\subfloat[case 1: without punctual outliers]{\includegraphics[scale=0.29]{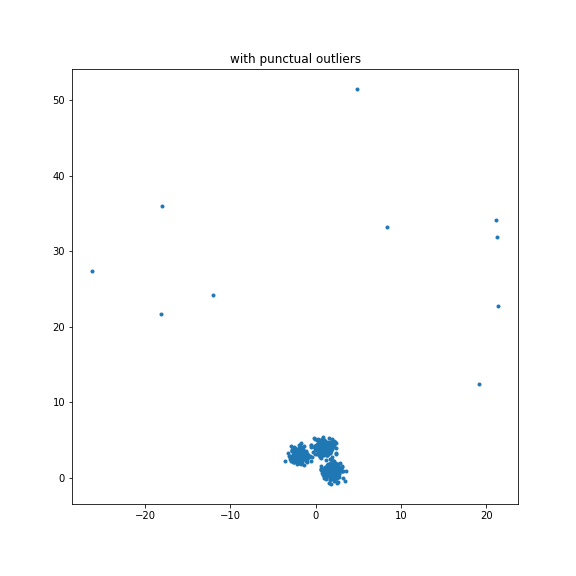}

}\subfloat[case 2: with a cluster of outliers]{\includegraphics[scale=0.29]{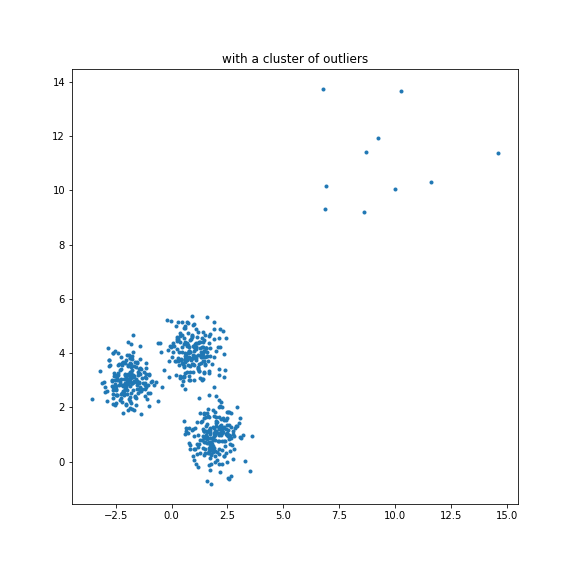}

}
\par\end{centering}
\caption{Illustrations of simulated data generated according to a Gaussian
Mixture Model in order to compare initialisation methods if the context
of outliers \label{fig:initialisation_datapoints}}
\end{figure}

The data are generated from $K=3$ multivariate Gaussian distributions
of dimension $p=2$ and length $n_{1}=n_{2}=n_{3}=300$ with variance
$\sigma^{2}=0.6$ and average vectors $\mu_{1}=[1,4],\mu_{2}=[2,1]$
and $\mu_{3}=[-2,3]$. Figure \ref{fig:initialisation_datapoints}.a
illustrates one realisation of the simulated context.
\begin{itemize}
\item \textbf{simulation 1}: \textit{punctual outliers}. From these $n=600$
datapoints, we randomly select $n_{outlier}$ as potential outliers
and their coordinates are multiplied by a constant term $\beta$ which
quantifies how far these outliers are from their own distribution.
We consider different level of pollution of data $n_{outlier}\in\{9,27\}$
and different degrees of outliers $\beta\in\{5,20\}$. Figure \ref{fig:initialisation_datapoints}.b
illustrates the data polluted by $n_{outlier}=9$ with degree $\beta=20$.
\item \textbf{simulation 2}: \textit{cluster of outliers. }A cluster of
outliers of size $n_{outlier}$ is generated according to a $2$-dimensional
Gaussian distribution with average $\mu_{outlier}=\beta[1,1]$ and
variance fixed to $\sigma^{2}=1$. Note that the size of the cluster
of outliers varies among $n_{outlier}\in\{9,27\}$ and the level distance
varies such that $\beta\in\{5,20\}$. Figure \ref{fig:initialisation_datapoints}.c
illustrates the cluster of outliers with $n_{outlier}=9$ and degree
$\beta=5$.
\end{itemize}
For all the methods, the number of clusters is supposed to be known
and fixed to $K=3$.

\paragraph{Initialisation strategies: }

We consider the following 3 traditional initialisation strategies:
\begin{itemize}
\item \textbf{Random initialisation}: we select K datapoints randomly and
without replacement as initial centers.
\item \textbf{kmeans++} proposed by \cite{Vassilvitskii07} which is maybe
the most widely used technique to initialise clustering algorithms.
The first center is taken from the data uniformly at random. Then
iteratively and until the number $K$ of chosen clusters is reached,
a new center is chosen from the datapoints with a probability which
increases exponentially with the distance $D^{2}(x,c)$ to the closest
centers already chosen. 
\item \textbf{kmedians++} is a variant of kmeans++. The same process is
iterated but the probability is computed with respect to $D(x,c)$
instead of $D^{2}(x,c)$. 
\end{itemize}
and a robust initialisation strategy developed by Hasan et al. in
2009 named ROBIN~\cite{al2009robust} which is a density-based approach:
\begin{itemize}
\item \textbf{ROBIN} (ROBust INitialisation) uses the Local Outlier Factor
approach (LOF)~\cite{breunig2000lof} to select, as initial centroids,
data points far away from each other and representative of dense regions
in the dataset. This approach requires to know the number of clusters
$K$ and the number of neighboring data points in order to compute
the LOF of each data point. In the experiment, the number of neighboring
datapoints has been fixed to 10. According to the chosen method, selected
datapoints changes drastically and it has to be noted that the best
approach is obtained for the approximation method where the algorithm
looks for the first LOF value that falls in $]1-\varepsilon,1+\varepsilon[$
. We chose this method and set $\varepsilon=0.2$.
\end{itemize}
The implementations that we used in this study for the above approaches
come from scikit-learn library which is a free software machine learning
library for the Python programming language and is publicly available~\cite{scikit-learn}.\\

We propose a robust variant of kmeans++ and kmedians++ by applying
the MOM principle as described in Section~\ref{sec:kB-MOM-algorithm}.
In particular, let $B$ be the number of blocks of data, $n_{B}$
the size of each block and $\mathcal{R}_{b}(c)$ the empirical risk
of the $b$th block. Then, we define the following algorithm :
\begin{enumerate}
\item Iterate from $1$ until $B$ blocks:
\begin{enumerate}
\item Select at random, uniformly and with replacement $n_{B}$ datapoints
\item Proceed a kmeans++ (or kmedians++) initialisation
\item Compute the empirical risk of the block $b$
\end{enumerate}
\item Select the centers from the block having the median empirical risk
\item Affect the datapoints to their nearest centroid of the selected (median)
block.
\end{enumerate}
Note that the size of each block is chosen equal to 18 and the number
of blocks is fixed to 250. These parameters follow the breakdown point
bounds presented in Section \ref{subsec:Breakdown-points}.

For the rest of the paper, we will call K-bMOM-km++ (respectively
K-bMOM-kmed++) the robust strategy based on K-means++ (respectively
kmedians++). 

\paragraph{Performance criteria:}

In order to compare the different starting strategies in terms of
performance, we compute 4 criteria:
\begin{itemize}
\item the Root Mean Square Error (RMSE) in order to evaluate the robustness
of fitted centers once the initialisation step is performed. This
criterion is calculated between the centers proposed by the initialisation
process and the ones used to simulate the data, given by:
\[
\text{RMSE}=\sqrt{\frac{\sum_{k=1}^{K}\left\Vert \hat{c}_{k}-\mu_{k}\right\Vert ^{2}}{K}}
\]
where $\hat{c}_{k}$ stands for the started center the most probable
for the class $k$ and $\mu_{k}$ the average parameter of the $k$th
mixture. 
\item the accuracy (acc) of the initial partition obtained by the nearest
initial centers and computed on the non-polluted data. This is equivalent
to a classification rate.
\item the Adjusted Rand Index (ARI) computed between the partition obtained
by the nearest initial centers and computed on the non-polluted data.
\item the empirical distortion obtained at the end of the initialisation
step and computed on the non polluted data:
\[
\hat{R}(\hat{\mathbf{c}})=\sum_{k=1}^{K}\sum_{x_{i}\in\mathcal{C}_{k}}\left\Vert x_{i}-\hat{c}_{k}\right\Vert ^{2}
\]
\item the number of clusters obtained on the non polluted data named \textit{nb}.
\end{itemize}
The experience has been repeated 300 times and for all these criteria,
average and standard deviations have been computed for each initialisation
method. 

\paragraph{Empirical Results for simulation 1: }

The results of simulation 1 are summarized in Table~\ref{tab:Table simulation1 init}. 

As we can observe, except for the random approach which behaves roughly
the same manner according to the different contexts, all the starting
approaches behave quite well when the number of outliers is small
($n_{outlier}\leq9$) and their distance level is low (cases $\beta=5$)
: accuracies vary between 0.92 to 0.98. However, ROBIN and the K-bMOM
based initialisation are the more stable approaches with a standard
deviation around 2 to 5\% whereas the 3 other methods remains up to
8.4\%. Besides, as soon as the context becomes harder (more outliers
and further), only the K-bMOM approaches have their accuracies and
ARIs unchanged whereas the performances of the 4 other methods decrease
drastically.

The level of the RMSE computed on the initial centers depends on the
strategy used: in particular, it remains under 1 in average for the
kmedians++, ROBIN, K-bMOM-km++ and K-bMOM-kmed++ strategies when the
simulated context is simple ($n_{outlier}=9$ , $\beta=5$). As the
distance level of outliers and the number of outliers increase, the
kmeans++ strategy propose poor centers since at least one of them
is stuck on an outlier. Indeed, its RMSE is up to 50 and the number
of clusters fitted on the non polluted data is below the true number
of components. The kmedians++ is more robust to outliers, by construction,
but its performances decrease drastically when both the number of
outliers and the distance level become higher ($n_{outlier}=27,\beta=20$).
The RMSE becomes up to 30 and the accuracy is about 0.77. At the opposite,
K-bMOM-km++ and K-bMOM-kmed++, well-perform in every contexts of simulations
even when the number of outliers reaches 27 and the distance level
20. In average, the initialisation by K-bMOM-km++ is 95 \% accurate
at the end of the initialisation step and the proposed centers remain
really close to theoritical ones (RMSE $<1$ in average). 

Finally, Figures \ref{fig:Boxplot simulation 1 init } and \ref{fig:Boxplot simulation 1 init - distoritions}
stand for violinplots of accuracies and distortions respectively for
each initialisation method from the less noisy simulation context
to the noisiest one. Several information are displayed in these violinplots:
the interquartile range (black bold vertical line), the median (orange
point), the percentile 95 (navy blue horizontal line) and the probability
density of accuracies (resp. distortions) for each method. In the
context $\left(n_{outlier},\beta\right)=\left(20,27\right)$, one
can observe the erratic behavior of ROBIN represented by the bimodal
distribution of its accuracy: it is true that in median this approach
reaches 95\% of accuracy but 10\% of the time, the initialisation
present poor results (under 60\% of accuracy) compared to K-bMOM which
does not decrease below 65\%. The same kind of observations can be
done on the distortions (see Figure \ref{fig:Boxplot simulation 1 init - distoritions}). 

Finally, by combining the results in distortions and accuracies K-bMOM-km++
and K-bMOM-kmed++ are the initialisation procedures which performs
the best in terms of stability and the accuracy of initial centers.
They are insensitive to the distance of outliers with the rest of
data and remain quite effective even when the number of outliers increases
(around 3\% of data in our context).

\begin{table}[p]
\begin{centering}
\begin{tabular}{|l|l|l|rrrrr|}
\hline 
\textsf{\footnotesize{}$n_{outlier}$} & \textsf{\footnotesize{}$\beta$} & \textsf{\textbf{\footnotesize{}Initialisation}} & \textsf{\textbf{\footnotesize{}RMSE}} & \textsf{\textbf{\footnotesize{}accuracy}} & \textsf{\textbf{\footnotesize{}ari}} & \textsf{\textbf{\footnotesize{}distortion}} & \textsf{\textbf{\footnotesize{}nb}}\tabularnewline
\hline 
\hline 
\textsf{\footnotesize{}9} & \textsf{\footnotesize{}5} & \textsf{\footnotesize{}random} & \textsf{\footnotesize{}1.738 (1.697)} & \textsf{\footnotesize{}0.763 (0.133)} & \textsf{\footnotesize{}0.564 (0.212)} & \textsf{\footnotesize{}3399.1 (1785.4)} & \textsf{\footnotesize{}3.0 (0.2)}\tabularnewline
\textsf{\footnotesize{}9} & \textsf{\footnotesize{}5} & \textsf{\footnotesize{}kmeans++} & \textsf{\footnotesize{}2.538 (3.598)} & \textsf{\footnotesize{}0.91 (0.13)} & \textsf{\footnotesize{}0.84 (0.187)} & \textsf{\footnotesize{}1559.1 (897.9)} & \textsf{\footnotesize{}2.8 (0.4)}\tabularnewline
\textsf{\footnotesize{}9} & \textsf{\footnotesize{}5} & \textsf{\footnotesize{}kmedians++} & \textsf{\footnotesize{}1.009 (1.365)} & \textsf{\footnotesize{}0.95 (0.084)} & \textsf{\footnotesize{}0.891 (0.141)} & \textsf{\footnotesize{}1306.6 (619.8)} & \textsf{\footnotesize{}3.0 (0.2)}\tabularnewline
\textsf{\footnotesize{}9} & \textsf{\footnotesize{}5} & \textsf{\footnotesize{}ROBIN} & \textsf{\footnotesize{}0.951 (0.45)} & \textsf{\textbf{\footnotesize{}0.973 (0.028)}} & \textsf{\footnotesize{}0.925 (0.063)} & \textsf{\footnotesize{}1385.0 (326.3)} & \textsf{\footnotesize{}3.0 (0.1)}\tabularnewline
\textsf{\footnotesize{}9} & \textsf{\footnotesize{}5} & \textsf{\footnotesize{}K-bMOM-km++} & \textsf{\textbf{\footnotesize{}0.457 (0.947)}} & \textsf{\textbf{\footnotesize{}0.988 (0.029)}} & \textsf{\textbf{\footnotesize{}0.968 (0.044)}} & \textsf{\textbf{\footnotesize{}790.0 (234.5)}} & \textsf{\textbf{\footnotesize{}3.0 (0.1)}}\tabularnewline
\textsf{\footnotesize{}9} & \textsf{\footnotesize{}5} & \textsf{\footnotesize{}K-bMOM-kmed++} & \textsf{\textbf{\footnotesize{}0.488 (0.815)}} & \textsf{\textbf{\footnotesize{}0.981 (0.053)}} & \textsf{\textbf{\footnotesize{}0.956 (0.088)}} & \textsf{\textbf{\footnotesize{}832.3 (342.3)}} & \textsf{\textbf{\footnotesize{}3.0 (0.1)}}\tabularnewline
\hline 
\textsf{\footnotesize{}9} & \textsf{\footnotesize{}20} & \textsf{\footnotesize{}random} & \textsf{\footnotesize{}2.432 (5.659)} & \textsf{\footnotesize{}0.771 (0.143)} & \textsf{\footnotesize{}0.58 (0.238)} & \textsf{\footnotesize{}3421.4 (2079.7)} & \textsf{\footnotesize{}3.0 (0.2)}\tabularnewline
\textsf{\footnotesize{}9} & \textsf{\footnotesize{}20} & \textsf{\footnotesize{}kmeans++} & \textsf{\footnotesize{}54.734 (10.795)} & \textsf{\footnotesize{}0.427 (0.147)} & \textsf{\footnotesize{}0.141 (0.226)} & \textsf{\footnotesize{}6807.2 (2869.2)} & \textsf{\footnotesize{}1.3 (0.5)}\tabularnewline
\textsf{\footnotesize{}9} & \textsf{\footnotesize{}20} & \textsf{\footnotesize{}kmedians++} & \textsf{\footnotesize{}7.884 (15.954)} & \textsf{\footnotesize{}0.907 (0.13)} & \textsf{\footnotesize{}0.835 (0.192)} & \textsf{\footnotesize{}1593.5 (952.0)} & \textsf{\footnotesize{}2.8 (0.4)}\tabularnewline
\textsf{\footnotesize{}9} & \textsf{\footnotesize{}20} & \textsf{\footnotesize{}ROBIN} & \textsf{\footnotesize{}1.317 (3.876)} & \textsf{\textbf{\footnotesize{}0.972 (0.037)}} & \textsf{\footnotesize{}0.924 (0.073)} & \textsf{\footnotesize{}1412.3 (376.2)} & \textsf{\footnotesize{}3.0 (0.1)}\tabularnewline
\textsf{\footnotesize{}9} & \textsf{\footnotesize{}20} & \textsf{\footnotesize{}K-bMOM-km++} & \textsf{\textbf{\footnotesize{}0.402 (0.162)}} & \textsf{\textbf{\footnotesize{}0.989 (0.009)}} & \textsf{\textbf{\footnotesize{}0.969 (0.026)}} & \textsf{\textbf{\footnotesize{}789.2 (150.7)}} & \textsf{\textbf{\footnotesize{}3.0 (0.0)}}\tabularnewline
\textsf{\footnotesize{}9} & \textsf{\footnotesize{}20} & \textsf{\footnotesize{}K-bMOM-kmed++} & \textsf{\textbf{\footnotesize{}0.393 (0.171)}} & \textsf{\textbf{\footnotesize{}0.987 (0.031)}} & \textsf{\textbf{\footnotesize{}0.966 (0.052)}} & \textsf{\textbf{\footnotesize{}801.5 (287.5)}} & \textsf{\textbf{\footnotesize{}3.0 (0.0)}}\tabularnewline
\hline 
\textsf{\footnotesize{}27} & \textsf{\footnotesize{}20} & \textsf{\footnotesize{}random} & \textsf{\footnotesize{}4.175 (9.975)} & \textsf{\footnotesize{}0.752 (0.143)} & \textsf{\footnotesize{}0.549 (0.229)} & \textsf{\footnotesize{}3506.7 (1891.5)} & \textsf{\footnotesize{}2.9 (0.3)}\tabularnewline
\textsf{\footnotesize{}27} & \textsf{\footnotesize{}20} & \textsf{\footnotesize{}kmeans++} & \textsf{\footnotesize{}57.84 (7.832)} & \textsf{\footnotesize{}0.343 (0.05)} & \textsf{\footnotesize{}0.012 (0.077)} & \textsf{\footnotesize{}8810.5 (2902.3)} & \textsf{\footnotesize{}1.0 (0.2)}\tabularnewline
\textsf{\footnotesize{}27} & \textsf{\footnotesize{}20} & \textsf{\footnotesize{}kmedians++} & \textsf{\footnotesize{}31.532 (19.748)} & \textsf{\footnotesize{}0.734 (0.156)} & \textsf{\footnotesize{}0.604 (0.222)} & \textsf{\footnotesize{}2782.4 (1378.7)} & \textsf{\footnotesize{}2.2 (0.5)}\tabularnewline
\textsf{\footnotesize{}27} & \textsf{\footnotesize{}20} & \textsf{\footnotesize{}ROBIN} & \textsf{\footnotesize{}25.71 (29.783)} & \textsf{\footnotesize{}0.738 (0.289)} & \textsf{\footnotesize{}0.585 (0.42)} & \textsf{\footnotesize{}4199.2 (3936.4)} & \textsf{\footnotesize{}2.3 (0.9)}\tabularnewline
\textsf{\footnotesize{}27} & \textsf{\footnotesize{}20} & \textsf{\footnotesize{}K-bMOM-km++} & \textsf{\textbf{\footnotesize{}3.361 (10.576)}} & \textsf{\textbf{\footnotesize{}0.951 (0.094)}} & \textsf{\textbf{\footnotesize{}0.903 (0.143)}} & \textsf{\textbf{\footnotesize{}1005.0 (507.3)}} & \textsf{\textbf{\footnotesize{}2.9 (0.3)}}\tabularnewline
\textsf{\footnotesize{}27} & \textsf{\footnotesize{}20} & \textsf{\footnotesize{}K-bMOM-kmed++} & \textsf{\textbf{\footnotesize{}4.786 (12.513)}} & \textsf{\textbf{\footnotesize{}0.934 (0.115)}} & \textsf{\textbf{\footnotesize{}0.882 (0.172)}} & \textsf{\textbf{\footnotesize{}1117.7 (677.9)}} & \textsf{\textbf{\footnotesize{}2.9 (0.3)}}\tabularnewline
\hline 
\end{tabular}
\par\end{centering}
\caption{{\footnotesize{}Average (and standard deviation in parentheses) of
accuracies and RMSE computed on 300 repetitions of the simulation
1 for the 6 proposed initialisation methods for different number of
outliers and distance levels.\label{tab:Table simulation1 init}}}
\end{table}
\begin{center}
\begin{figure}[p]
\begin{centering}
\includegraphics[scale=0.42]{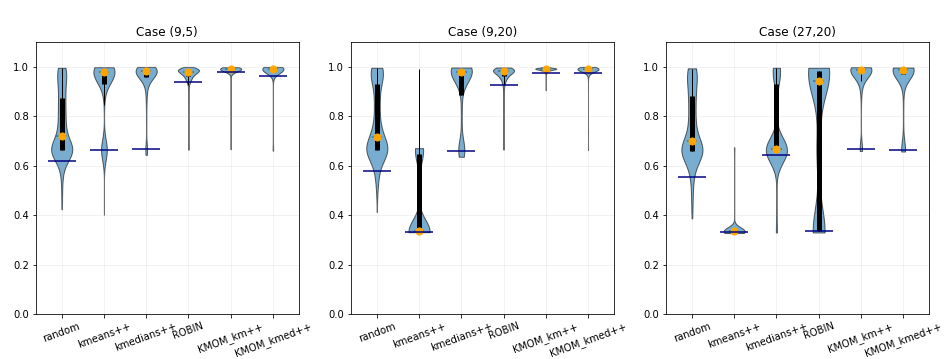}
\par\end{centering}
\begin{centering}
\caption{{\footnotesize{}Violinplots of accuracies of 6 initialisation approaches
according to the level of pollution of data in the context of punctual
outliers. From the less noisy context (left) to the noisiest one (right).\label{fig:Boxplot simulation 1 init }}}
\par\end{centering}
\begin{centering}
\includegraphics[scale=0.42]{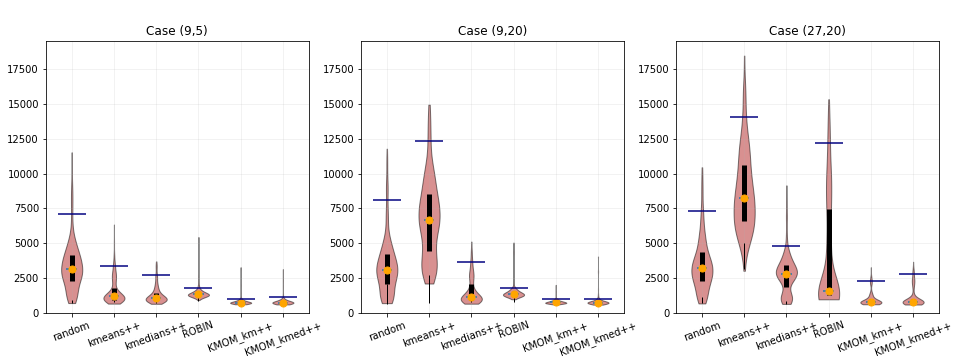}
\par\end{centering}
\caption{{\footnotesize{}Violinplots of distortions of 6 initialisation approaches
according to the level of pollution of data in the context of punctual
outliers. From the less noisy context (left) to the noisiest one (right).\label{fig:Boxplot simulation 1 init - distoritions}}}
\end{figure}
\par\end{center}

\paragraph{Empirical Results for simulation 2: }

The results of simulation 2 are summarized in Table \ref{tab:Table simulation 2 init}. 

\begin{figure}[ph]
\begin{centering}
\includegraphics[scale=0.41]{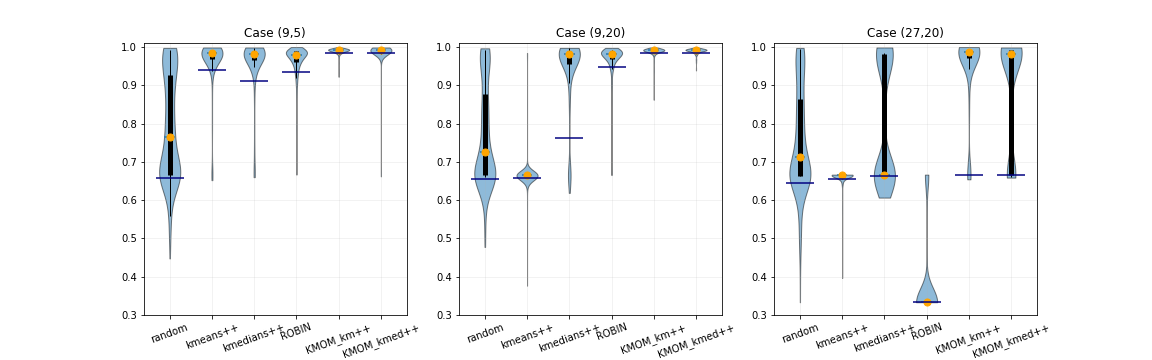}
\par\end{centering}
\caption{{\footnotesize{}Violinplots of accuracies of 6 initialisation approaches
according to the level of pollution of data in the context of cluster
of outliers. From the less noisy context (left) to the noisiest one
(right).\label{fig:Boxplot simulation 2 init}}}

\begin{centering}
\includegraphics[scale=0.41]{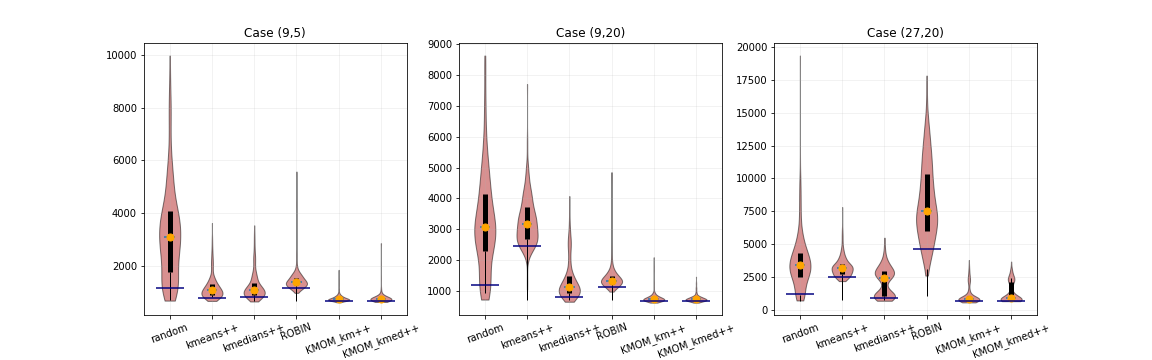}
\par\end{centering}
\caption{{\footnotesize{}Violinplots of distortions of 6 initialisation approaches
according to the level of pollution of data in the context of cluster
of outliers. From the less noisy context (left) to the noisiest one
(right).\label{fig:Boxplot simulation 2 - distortions}}}
\end{figure}

Again, in this situation the random initialization is not as bad as
we could expect in average, however such a starting approach is very
instable as we can observe via its standard deviations. On the other
hand, the standard initialization methods based on kmeans++ and kmedians++
(at least in accuracy) present comparable performances to their robust
version for a low number of outliers (see case $n_{outlier}=9$ for
$\beta=5$). This can be explained simply by the fact that the outliers
are grouped together in the same area of the space and therefore kmeans++
and kmedians++ are going to chose started centers well-spread among
the datasets by construction. However, when the number of outliers
increases and so does their distance to the grouped data, then they
are outperformed by their robust versions. Finally, Figure \ref{fig:Boxplot simulation 2 init}
stands for boxplots of all accuracies (left) and all RMSE (right)
over the noisiest versions of the simulation context of a cluster
of outliers which groups together $n_{outlier}=27$ and $\beta=\{5,20\}$.
Again, the K-bMOM-km++ initialisation presents better and stable results
in both accuracy and RMSE compared to the rest of methods. 

\begin{table}[ph]
\begin{centering}
\begin{tabular}{|lll|rrrrr|}
\hline 
\textsf{\footnotesize{}$n_{outlier}$} & \textsf{\footnotesize{}$\beta$} & \multicolumn{1}{l}{\textsf{\textbf{\footnotesize{}Initialisation}}} & \textsf{\textbf{\footnotesize{}RMSE}} & \textsf{\textbf{\footnotesize{}accuracy}} & \textsf{\textbf{\footnotesize{}ari}} & \textsf{\textbf{\footnotesize{}distortion}} & \textsf{\textbf{\footnotesize{}nb}}\tabularnewline
\hline 
\hline 
\textsf{\footnotesize{}9} & \textsf{\footnotesize{}5} & \textsf{\footnotesize{}random} & \textsf{\scriptsize{}1.429 (0.663)} & \textsf{\scriptsize{}0.791 (0.138)} & \textsf{\scriptsize{}0.609 (0.226)} & \textsf{\scriptsize{}3239.7 (1795.4)} & \textsf{\scriptsize{}3.0 (0.1)}\tabularnewline
\textsf{\footnotesize{}9} & \textsf{\footnotesize{}5} & \textsf{\footnotesize{}kmeans++} & \textsf{\scriptsize{}0.743 (0.307)} & \textsf{\scriptsize{}0.962 (0.07)} & \textsf{\scriptsize{}0.912 (0.122)} & \textsf{\scriptsize{}1193.9 (464.1)} & \textsf{\scriptsize{}3.0 (0.1)}\tabularnewline
\textsf{\footnotesize{}9} & \textsf{\footnotesize{}5} & \textsf{\footnotesize{}kmedians++} & \textsf{\scriptsize{}0.777 (0.347)} & \textsf{\scriptsize{}0.955 (0.077)} & \textsf{\scriptsize{}0.896 (0.137)} & \textsf{\scriptsize{}1239.6 (517.0)} & \textsf{\scriptsize{}3.0 (0.1)}\tabularnewline
\textsf{\footnotesize{}9} & \textsf{\footnotesize{}5} & \textsf{\footnotesize{}ROBIN} & \textsf{\scriptsize{}0.948 (0.161)} & \textsf{\scriptsize{}0.97 (0.032)} & \textsf{\scriptsize{}0.916 (0.074)} & \textsf{\scriptsize{}1437.3 (369.1)} & \textsf{\scriptsize{}3.0 (0.1)}\tabularnewline
\textsf{\footnotesize{}9} & \textsf{\footnotesize{}5} & \textsf{\footnotesize{}K-bMOM-km++} & \textsf{\textbf{\scriptsize{}0.368 (0.141)}} & \textsf{\textbf{\scriptsize{}0.99 (0.008)}} & \textsf{\textbf{\scriptsize{}0.971 (0.023)}} & \textsf{\textbf{\scriptsize{}772.6 (125.5)}} & \textsf{\textbf{\scriptsize{}3.0 (0.0)}}\tabularnewline
\textsf{\footnotesize{}9} & \textsf{\footnotesize{}5} & \textsf{\footnotesize{}K-bMOM-kmed++} & \textsf{\textbf{\scriptsize{}0.376 (0.197)}} & \textsf{\textbf{\scriptsize{}0.987 (0.034)}} & \textsf{\textbf{\scriptsize{}0.965 (0.06)}} & \textsf{\textbf{\scriptsize{}790.8 (244.6)}} & \textsf{\textbf{\scriptsize{}3.0 (0.0)}}\tabularnewline
\hline 
\textsf{\footnotesize{}9} & \textsf{\footnotesize{}20} & \textsf{\footnotesize{}ranom} & \textsf{\scriptsize{}1.4 (0.608)} & \textsf{\scriptsize{}0.771 (0.131)} & \textsf{\scriptsize{}0.582 (0.211)} & \textsf{\scriptsize{}3220.1 (1577.0)} & \textsf{\scriptsize{}3.0 (0.2)}\tabularnewline
\textsf{\footnotesize{}9} & \textsf{\footnotesize{}20} & \textsf{\footnotesize{}kmeans++} & \textsf{\scriptsize{}1.058 (0.608)} & \textsf{\scriptsize{}0.666 (0.039)} & \textsf{\scriptsize{}0.513 (0.074)} & \textsf{\scriptsize{}3280.9 (779.8)} & \textsf{\scriptsize{}2.0 (0.1)}\tabularnewline
\textsf{\footnotesize{}9} & \textsf{\footnotesize{}20} & \textsf{\footnotesize{}kmedians++} & \textsf{\scriptsize{}0.795 (0.32)} & \textsf{\scriptsize{}0.94 (0.098)} & \textsf{\scriptsize{}0.877 (0.152)} & \textsf{\scriptsize{}1359.4 (648.8)} & \textsf{\scriptsize{}2.9 (0.3)}\tabularnewline
\textsf{\footnotesize{}9} & \textsf{\footnotesize{}20} & \textsf{\footnotesize{}ROBIN} & \textsf{\scriptsize{}0.921 (0.132)} & \textsf{\textbf{\scriptsize{}0.974 (0.032)}} & \textsf{\textbf{\scriptsize{}0.928 (0.066)}} & \textsf{\scriptsize{}1401.6 (348.9)} & \textsf{\textbf{\scriptsize{}3.0 (0.1)}}\tabularnewline
\textsf{\footnotesize{}9} & \textsf{\footnotesize{}20} & \textsf{\footnotesize{}K-bMOM-km++} & \textsf{\textbf{\scriptsize{}0.37 (0.147)}} & \textsf{\textbf{\scriptsize{}0.989 (0.011)}} & \textsf{\textbf{\scriptsize{}0.969 (0.029)}} & \textsf{\textbf{\scriptsize{}772.3 (141.8)}} & \textsf{\textbf{\scriptsize{}3.0 (0.0)}}\tabularnewline
\textsf{\footnotesize{}9} & \textsf{\footnotesize{}20} & \textsf{\footnotesize{}K-bMOM-kmed++} & \textsf{\textbf{\scriptsize{}0.359 (0.132)}} & \textsf{\textbf{\scriptsize{}0.99 (0.007)}} & \textsf{\textbf{\scriptsize{}0.971 (0.02)}} & \textsf{\textbf{\scriptsize{}763.5 (106.7)}} & \textsf{\textbf{\scriptsize{}3.0 (0.0)}}\tabularnewline
\hline 
\textsf{\footnotesize{}27} & \textsf{\footnotesize{}20} & \textsf{\footnotesize{}random} & \textsf{\scriptsize{}1.455 (0.705)} & \textsf{\scriptsize{}0.755 (0.137)} & \textsf{\scriptsize{}0.552 (0.22)} & \textsf{\scriptsize{}3656.2 (2096.8)} & \textsf{\scriptsize{}2.9 (0.3)}\tabularnewline
\textsf{\footnotesize{}27} & \textsf{\footnotesize{}20} & \textsf{\footnotesize{}kmeans++} & \textsf{\scriptsize{}0.962 (0.552)} & \textsf{\scriptsize{}0.661 (0.019)} & \textsf{\scriptsize{}0.506 (0.059)} & \textsf{\scriptsize{}3264.6 (756.5)} & \textsf{\scriptsize{}2.0 (0.0)}\tabularnewline
\textsf{\footnotesize{}27} & \textsf{\footnotesize{}20} & \textsf{\footnotesize{}kmedians++} & \textsf{\scriptsize{}0.925 (0.494)} & \textsf{\scriptsize{}0.807 (0.156)} & \textsf{\scriptsize{}0.707 (0.214)} & \textsf{\scriptsize{}2179.2 (1084.5)} & \textsf{\scriptsize{}2.5 (0.5)}\tabularnewline
\textsf{\footnotesize{}27} & \textsf{\footnotesize{}20} & \textsf{\footnotesize{}ROBIN} & \textsf{\scriptsize{}2.036 (1.295)} & \textsf{\scriptsize{}0.38 (0.115)} & \textsf{\scriptsize{}0.068 (0.17)} & \textsf{\scriptsize{}8219.6 (2931.6)} & \textsf{\scriptsize{}1.1 (0.3)}\tabularnewline
\textsf{\footnotesize{}27} & \textsf{\footnotesize{}20} & \textsf{\footnotesize{}K-bMOM-km++} & \textsf{\textbf{\scriptsize{}0.548 (0.354)}} & \textsf{\textbf{\scriptsize{}0.94 (0.108)}} & \textsf{\textbf{\scriptsize{}0.89 (0.161)}} & \textsf{\textbf{\scriptsize{}1106.5 (652.8)}} & \textsf{\textbf{\scriptsize{}2.9 (0.3)}}\tabularnewline
\textsf{\footnotesize{}27} & \textsf{\footnotesize{}20} & \textsf{\footnotesize{}K-bMOM-kmed++} & \textsf{\scriptsize{}0.658 (0.429)} & \textsf{\scriptsize{}0.893 (0.141)} & \textsf{\scriptsize{}0.821 (0.207)} & \textsf{\textbf{\scriptsize{}1329.6 (762.6)}} & \textsf{\scriptsize{}2.8 (0.4)}\tabularnewline
\hline 
\end{tabular}
\par\end{centering}
\caption{{\footnotesize{}Average (and standard deviation in parentheses) of
RMSE, accuracies, distortions and number of clusters computed on 300
repetitions of the simulation 2 for the 6 proposed initialisation
methods for different number of outliers in the cluster of outliers
and different distance levels.\label{tab:Table simulation 2 init}}}
\end{table}

\paragraph{Conclusion:}

We showed in this Section that it seems therefore preferable to use
the robust version of popular initialization methods in the context
of outliers. In regards to the easiest context (small number of outliers
which are relatively close to the sample), where the traditionnal
kmeans++ initialisation works well and so its robust version, we could
recommand to the practioner to use all the time its robust version
even without outliers as robust initialisation in order to avoid the
sensitivity of clustering algorithm to initialization. An other asset
of such an initialisation process is the decrease of the computational
cost: on one hand the subsampling strategy itself enables to restrict
the space to be covered ; on a second hand, this strategy applied
on blocks independantly can be easily and highly parallelized. This
should be very benefic for large datasets. Besides, these both aspects
are going to be developed in Section~\ref{sec:Color-quantization}
on an application on color quantization on image processing. 

\subsection{Guidelines for the selection of hyperparameters linked to blocks}

The good behavior of our algorithm with respect to outliers is linked
to an appropriate choice of the size of blocks $n_{B}$ and the number
of blocks $B$. For a known level of noise, we are able to compute
lower and upper bounds respectively for the within-block size and
the number of blocks as presented in Section \ref{subsec:Breakdown-points},
enabling therefore to guide the practitioner. However, when the number
of outlier is unknown, it is important to propose a heuristic which
selects automatically the size of the blocks $n_{B}$. 

The proposed strategy is the following: the within block size varies
a priori from $K$ to $n/K$ and for each level of within block size,
the empirical risk of each block is computed and the median one is
kept and plotted. We choose $n_{B}^{*}$ the level of the size block
linked to a cutting-point of the curve. Indeed, as the within block
size increases the probability of picking an outlier in the block
and among all $B$ blocks increases and this should drastically impact
the empirical risk of the median block, hence the search of breakpoints
in this empirical risk. \\

In order to illustrate such a strategy, we consider a $2$-dimensional
Gaussian mixture models of $K=3$ components with equal size $n_{1}=n_{2}=n_{3}=300$.
The mean vectors are set to $\mu_{1}=[3,12],\mu_{2}=[6,3]$ and $\mu_{3}=[-6,9]$
and the variance parameter is set to $\sigma^{2}=0.6$. Twenty outliers
are selected randomly from the data and their coordinates are multiplied
by $50$. We look for 2 situations where we fix the number of blocks
to $B=50$ and $B=100$. 

\begin{figure}
\begin{centering}
\subfloat[case $B=50$ blocks]{\begin{centering}
\includegraphics[scale=0.2]{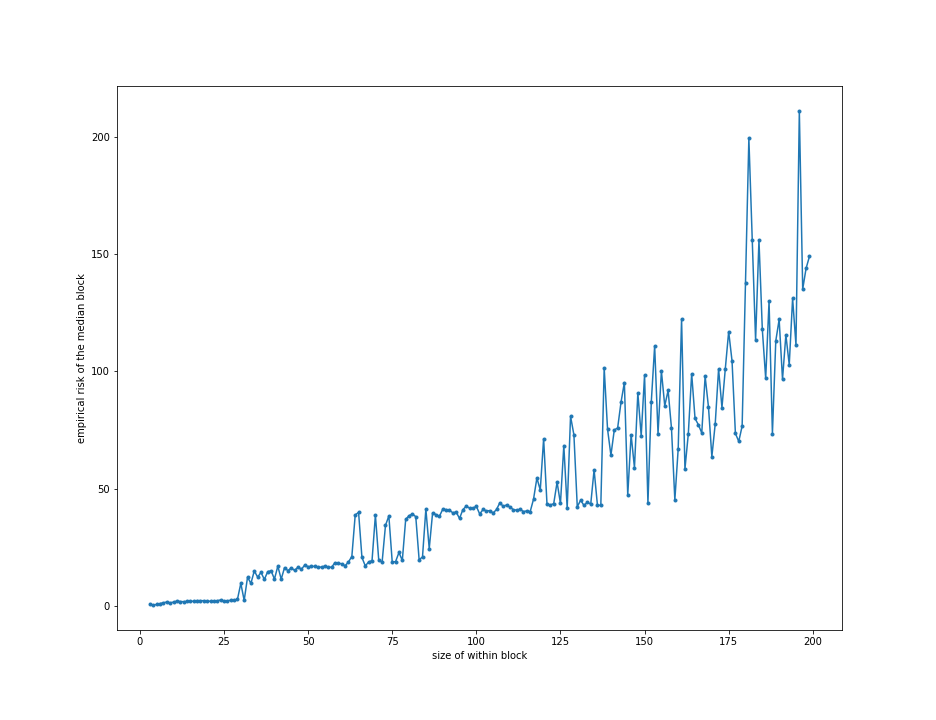}
\par\end{centering}
}\subfloat[case $B=100$ blocks]{\begin{centering}
\includegraphics[scale=0.2]{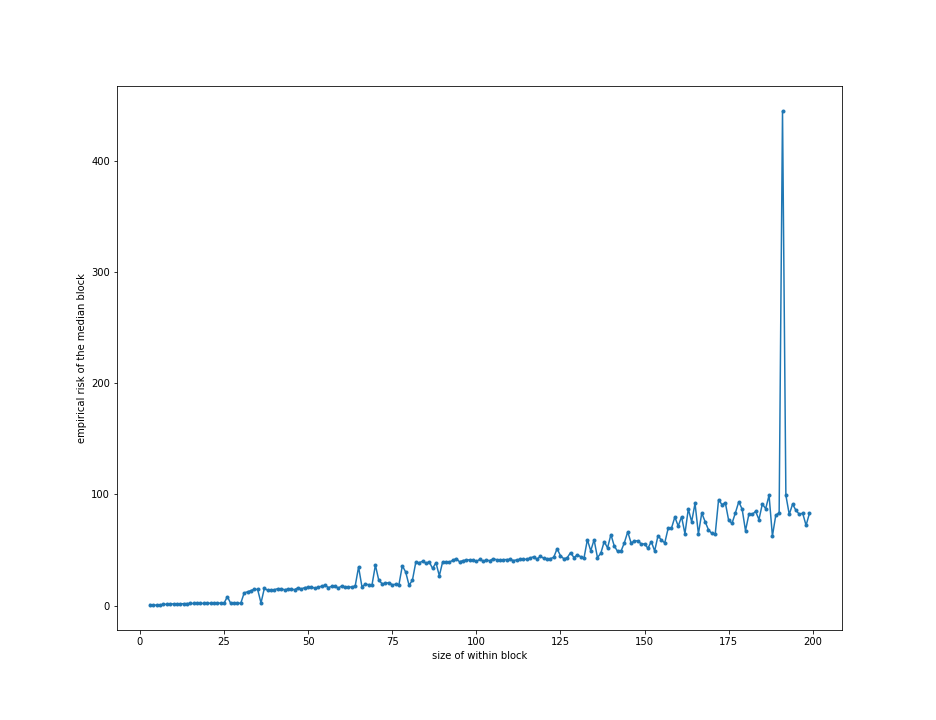}
\par\end{centering}
}
\par\end{centering}
\caption{Evolution of the empirical risk of the median block for $B=50$ blocks
(left) and $B=100$ blocks (right)\label{fig:Evolution_emprisk_hyperprms_calibration}}

\begin{centering}
\subfloat[case $B=50$ blocks]{\begin{centering}
\includegraphics[scale=0.2]{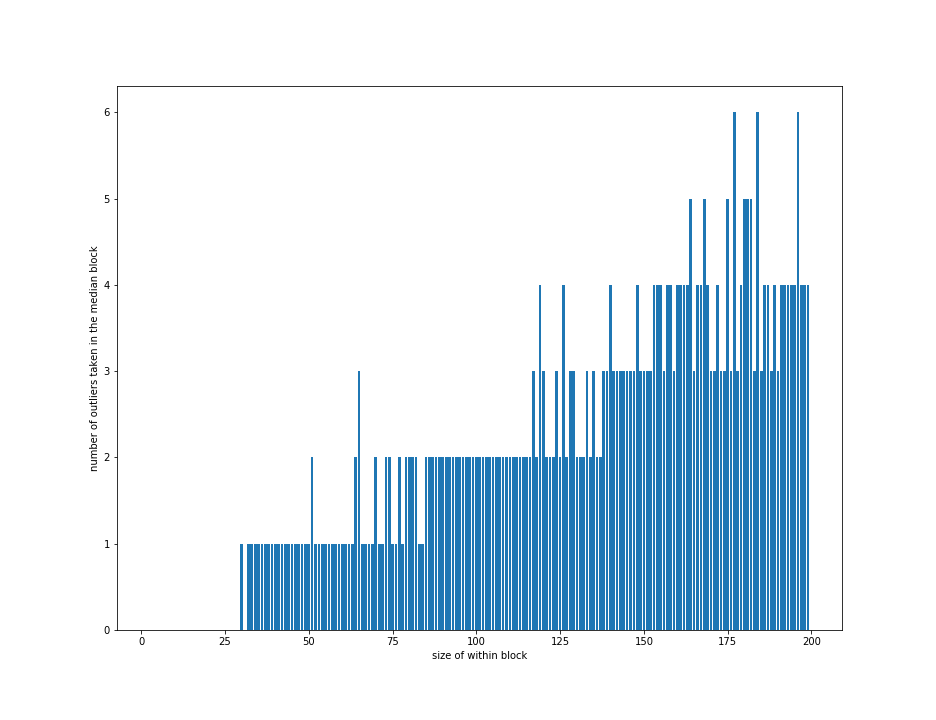}
\par\end{centering}
}\subfloat[case $B=100$ blocks]{\begin{centering}
\includegraphics[scale=0.2]{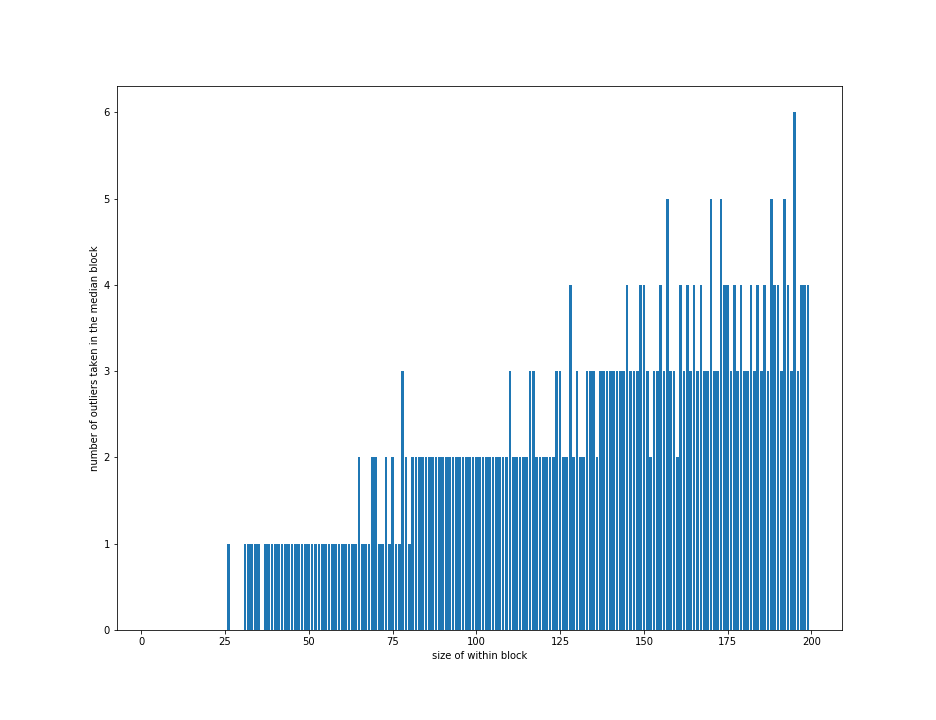}
\par\end{centering}
}
\par\end{centering}
\caption{Evolution of the number of outliers selected in the median block for
$B=50$ blocks (left) and $B=100$ blocks (right)\label{fig:Evolution_nb_outliers_hyperprms_calibration}}

\begin{centering}
\subfloat[case $B=50$ blocks]{\begin{centering}
\includegraphics[scale=0.2]{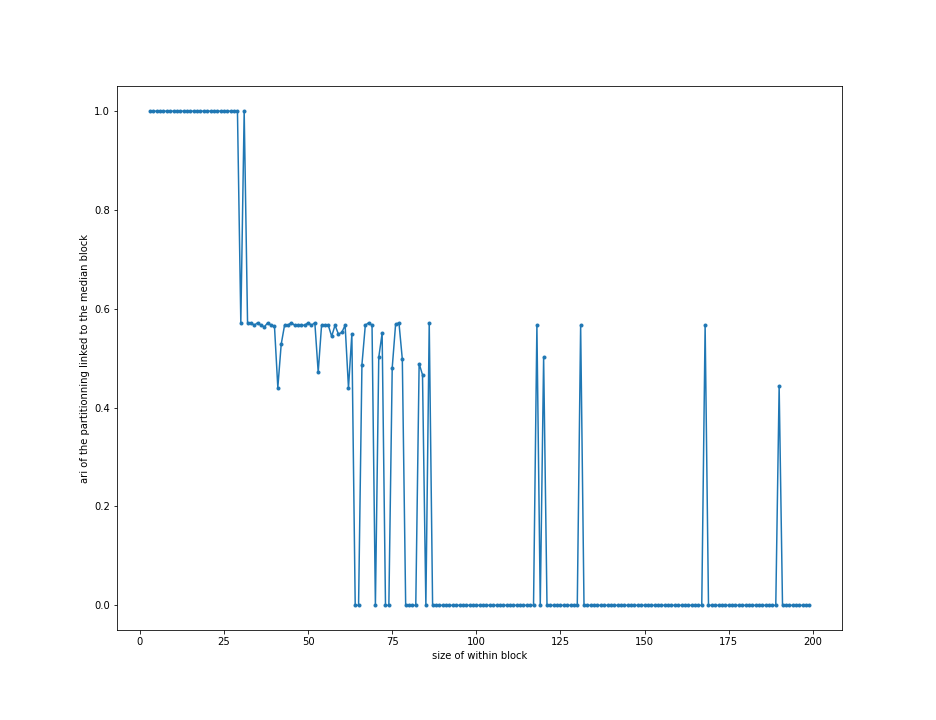}
\par\end{centering}
}\subfloat[case $B=100$ blocks]{\begin{centering}
\includegraphics[scale=0.2]{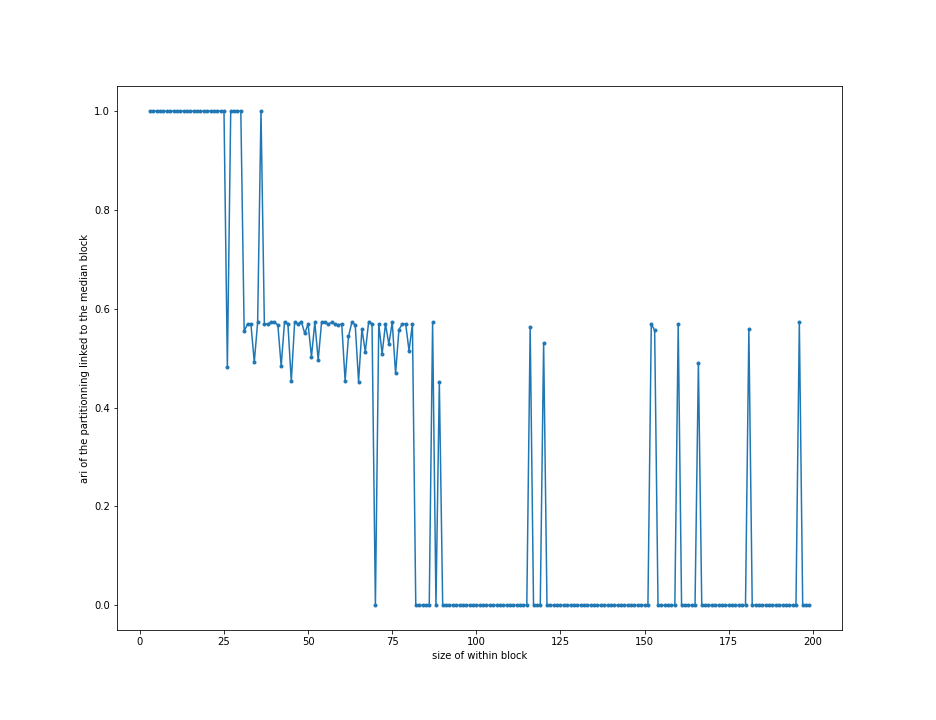}
\par\end{centering}
}
\par\end{centering}
\caption{Evolution of the ARI obtained by the partionning associated the median
block for $B=50$ blocks (left) and $B=100$ blocks (right)\label{fig:Evolution_ari_hyperprms_calibration}}
\end{figure}

Figure~\ref{fig:Evolution_emprisk_hyperprms_calibration}, Figure~\ref{fig:Evolution_nb_outliers_hyperprms_calibration}
and Figure~\ref{fig:Evolution_ari_hyperprms_calibration} depict
respectively the evolution of the median empirical risk, the number
of outliers present in the median block and the Adjusted Rand Index
(ARI) computed on the partitionning of data obtained by the nearest
centroid selected in the median block, according to the number of
data in the blocks.

We get $n_{B}^{*}\leq25$ for both cases as we can observe the evolution
of the empirical risk of the median block in Figure \ref{fig:Evolution_emprisk_hyperprms_calibration}a.
for the case with a number of blocks $B=50$ and in Figure \ref{fig:Evolution_emprisk_hyperprms_calibration}b.
for the case $B=100$. 

Note that the selection of $n_{B}^{*}$ works well in both examples
and the associated clustering seems also good. Indeed, under the selected
$n_{B=50}^{*}=20$ and $n_{B=100}^{*}=25$, there is no outlier present
in the median block and the resulting partitionning of data is perfect
on the non polluted data (ARI = 1). Above this cutting-point, the
number of outliers in the median block increases with the within block
size whereas the ARI index decreases.

These results show that, in practice, if one chooses a small size
of blocks and a high number of blocks, then the initialisation step
is likely to be robust. 

\subsection{Benchmark among the robust K-means-based algorithms}

\begin{figure}
\begin{centering}
\subfloat[Data without outlier\label{fig:Data-without-outlier}]{\includegraphics[scale=0.35]{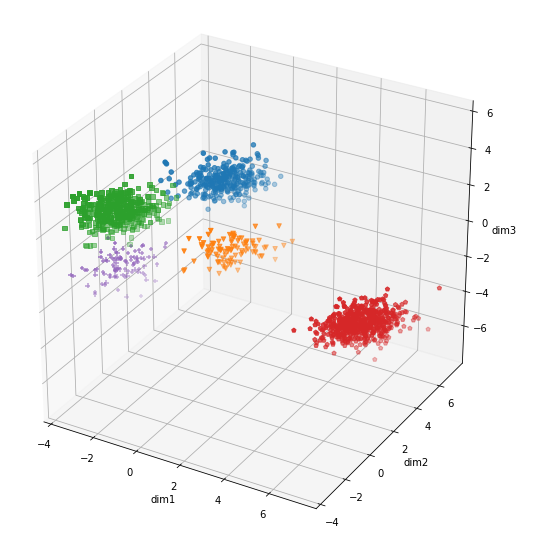}

}\hspace{2ex}\subfloat[Data with outliers\label{fig:Data-with-outliers}]{\includegraphics[scale=0.35]{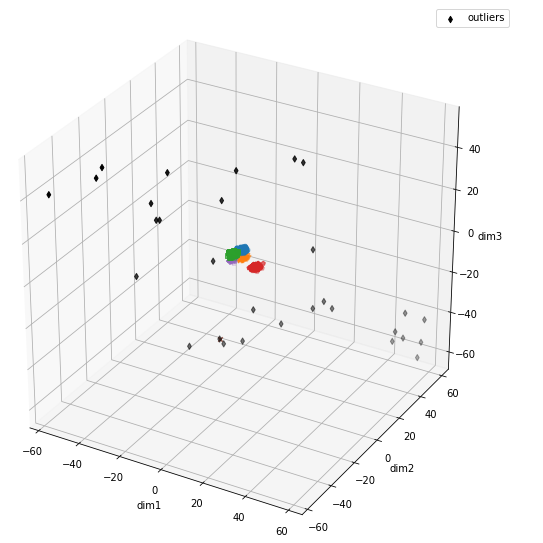}

}
\par\end{centering}
\caption{Illustration of generated data. }

\end{figure}

The objective of that section is to compare the performance of the
K-bMOM strategy with the robust clustering algorithms based on K-means
approaches on a framework with outliers. To do so, we dispose of $N=1500$
points of dimension $p=3$ which are generated according to a mixture
of $K=5$ multivariate Gaussian density functions with isotropic covariance
matrix. The average vectors for the 5 components are respectively
$\mu_{1}=[0,1,4]$, $\mu_{2}=[2,1,0]$, $\mu_{3}=[0,-2,3]$, $\mu_{4}=[0,5,-5]$
and $\mu_{5}=[-1,-2,0]$. An example of data generated according to
this framework is displayed in Figure~\ref{fig:Data-without-outlier}.
Outliers have been generated by randomly taken $30$ datapoints from
which their coordinates have been multiplied by a factor of +/-10.
An example of the final polluted data are illustrated in Figure~\ref{fig:Data-with-outliers}.

Given this context, three variations from this framework have been
considered in this Section:
\begin{description}
\item [{Variation~1}] The clusters have equal size and dispose of the
same spherical covariance matrix. These assumptions are well-suited
for the K-means procedure.
\item [{Variation~2}] The clusters have unequal size but dispose of the
same spherical covariance matrix.
\item [{Variation~3}] The clusters have unequal size and dispose of different
scaling parameters.
\end{description}
Simulation parameters for each of these variations are detailed below:
\begin{center}
\begin{tabular}{lll}
\hline 
\textsf{\small{}Variation} & \textsf{\small{}Size} & \textsf{\small{}Scaling parameter : $\sigma_{k}^{2}$}\tabularnewline
\hline 
\textsf{\small{}1} & \textsf{\small{}$\forall k\in\{1,\dots,5\}:n_{k}=n=300$} & \textsf{\small{}$\forall k\in\{1,\dots,5\}:\sigma_{k}^{2}=\sigma^{2}=0.6$}\tabularnewline
\textsf{\small{}2} & \textsf{\small{}$n_{1}=300,n_{2}=n_{5}=100,n_{3}=400,n_{4}=600$} & \textsf{\small{}$\forall k\in\{1,\dots,5\}:\sigma_{k}^{2}=\sigma^{2}=0.6$}\tabularnewline
\textsf{\small{}3} & \textsf{\small{}$n_{1}=300,n_{2}=n_{5}=100,n_{3}=400,n_{4}=600$} & \textsf{\small{}$\sigma_{1}^{2}=\sigma_{4}^{2}=1,\sigma_{2}^{2}=0.4,\sigma_{3}^{2}=0.6,\sigma_{5}^{2}=0.5$}\tabularnewline
\hline 
\end{tabular}
\par\end{center}

\vspace{2ex}

These variations have been repeated 50 times and each time, the K-means-based
algorithms have been initialized in the same manner with a K-means++
procedure iterated 10 times. \\

We consider 5 different algorithms : our proposed robust clustering
algorithm named K-bMOM and also 4 well-known robust versions of the
K-means. These methods are described below:
\begin{lyxlist}{00.00.0000}
\item [{\textbf{K-bMOM}}] K-bMOM algorithm introduced in Section~\ref{sec:kB-MOM-algorithm}.
\item [{\textbf{K-medoids}}] aims at finding $K$ data points as centers
such as the within inertia is minimized. The partition around medoids
algorithm named PAM~\cite{rdusseeun1987clustering} aims to achieve
this in two steps~: an assignement step where each datapoint is assigned
to its closest medoid; a refinement step which looks for better medoids
than the current ones. The search is each time exhaustive in the data
PAM has a complexity dominated by $\mathcal{O}\left(n^{2}kp\right)$.
Faster versions have been proposed in~\cite{schubert2018faster}.
The number of clusters $K$ needs to be set in the procedure.
\item [{\textbf{K-medians}}] is a robust variant of the $k$-means algorithm~\cite{jain1988algorithms}
: in the aggregation step, instead of computing the barycenter of
each group as in the $K$-means procedure, the $K$-medians compute
in each single dimension the median in the Manhattan-distance formulation.
This makes the algorithm more reliable for extreme values. The number
of clusters $K$ needs to be specified by the practitioner.
\item [{\textbf{trimmed-K-means}}] (trim-km) implementation is an EM-like
algorithm introduced by~\cite{cuesta1997trimmed} in the late 90s.
It is derived from the $K$-means and benefits robustness properties
from the trimming action during the maximisation step where only a
proportion $1-\alpha$ of the closest data point from their assigned
centroid is taken into account. Since the trimming needs to sort the
data points according to their distance to centroid, it leads therefore
to an overall complexity of $\mathcal{O}\left(nkp+n.\log n\right)$
at each iteration. Besides, note that in practice, the user needs
to choose a value $\alpha$ for the proportion of datapoints to be
discarded and no practical information is given to calibrate such
an hyperparameter. In the simulations, $\alpha$ is set to the true
value of the number of outliers ie $n_{outlier}$. 
\item [{\textbf{K-PDTM}}] is a robust quantization algorithm introduced
by~\cite{brecheteau2018robust} that aims to infer the manifold from
which the data points are drawn. This inference is done by means of
$K$ centroids that should be on the manifold if the algorithm runs
well. It is based also on a Lloyd-type algorithm where in the updating
step, the centroid is computed as the barycenter of the $q$ nearest
neighbours of the barycenter of the cluster. In the assignement step,
the data point is assigned according to a Bregman divergence. This
algorithm has two hyperparameters: $q$, the number of neighbors used
to compute the centroid and the number of clusters $K$.
\end{lyxlist}
Finally, by default, for all the proposed methods having the number
of clusters as hyperparameter, we set it to its true value ie $K=5$.
Moreover, 2 types of initialisations have been done: on one hand,
on the 3 first experiments the robust kmeans-based methods have been
initialized with a kmeans++ initialisation and on a second hand, these
algorithms have been initialized with the K-bMOM-km++ presented in
Section~\ref{subsec:Comparing-initialisation-strategies} on the
first context of simulation.

The implementations used for the clustering approaches to compare
the MOM-based ones in this experiment are publicly available. Table~\ref{tab:Implementations-and-hyperprms}
details the programming languages and associated librairies used as
well as selected hyperparameters. 

\begin{table}
\begin{centering}
\begin{tabular}{lll}
\hline 
\textsf{\small{}Algorithm} & \textsf{\small{}Language} & \textsf{\small{}Hyperparameters}\tabularnewline
\hline 
 &  & \tabularnewline
\textsf{\small{}k-means} & \textsf{\small{}Python}~\cite{scikit-learn} & \textsf{\small{}$K=5$, $init=$initial\_centers{*},$n\text{\_}init=1$}\tabularnewline
 &  & \tabularnewline
\textsf{\small{}k-medoids} & \textsf{\small{}Python}~\textsf{\small{}\cite{Novikov2019}} & \textsf{\small{}$initial\text{\_}index\text{\_}medoids=$ index of
the nearest datapoints of initial\_centers{*}}\tabularnewline
 &  & \tabularnewline
\textsf{\small{}k-medians} & \textsf{\small{}Python}~\textsf{\small{}\cite{Novikov2019}} & \textsf{\small{}$initial\text{\_}centers=$initial\_centers{*}}\tabularnewline
 &  & \tabularnewline
\textsf{\small{}trimmed k-means} & \textsf{\small{}R}~\textsf{\small{}\cite{TrimKmeans12}} & \textsf{\small{}$K=5$, $trim=$nb\_outliers/$N$, $runs=50$, $points=$initial\_centers{*},$maxit=5$}\tabularnewline
 &  & \tabularnewline
\textsf{\small{}k-pdtm} & \textsf{\small{}Python$^{1}$\cite{brecheteau2018robust}} & \textsf{\small{}$K=5$, $query\text{\_}pts=$initial\_centers{*},$q=5$,
$k=5$, $sig=N-$nb\_outliers,}\tabularnewline
 &  & \textsf{\small{}$iter\text{\_}max=300$, $nstart=1$,$leaf\text{\_}size=30$}\tabularnewline
 &  & \tabularnewline
\textsf{\small{}k-bmom} & \textsf{\small{}Python$^{2}$} & \textsf{\small{}$K=5$, $n_{B}=20$, $B=500$, $iter\text{\_}max=50$}\tabularnewline
 &  & \tabularnewline
\hline 
\multicolumn{3}{l}{\textsf{\scriptsize{}{*}initial\_centers : same centers obtained either
with a kmeans++ initialisation or kmom-km++}}\tabularnewline
\multicolumn{3}{l}{$^{1}$\textsf{\scriptsize{}\href{https://www.math.sciences.univ-nantes.fr/~brecheteau/notebooks/Notebook_kPDTM_kPLM.html}{https://www.math.sciences.univ-nantes.fr/$\sim$brecheteau/notebooks/Notebook\_kPDTM\_kPLM.html}}}\tabularnewline
\multicolumn{3}{l}{$^{2}$\textsf{\scriptsize{}\href{https://github.com/csaumard/kbMOM}{https://github.com/csaumard/kbMOM}}}\tabularnewline
\end{tabular}
\par\end{centering}
\caption{Implementations and hyperparameters\label{tab:Implementations-and-hyperprms}}
\end{table}
In order to compare the performances of these algorithms, the distortion
and the Adjusted Rand Index (ARI) have been computed based on the
true parameters of data distribution and their label membership. Moreover,
the average number of clusters found among the non polluted data have
also been displayed. 

\subsection*{Results and Analysis}

\begin{figure}[p]
\begin{centering}
\subfloat[Violinplots of ARI\label{fig:Benchmark_Violinplots_ARI}]{\begin{centering}
\includegraphics[scale=0.41]{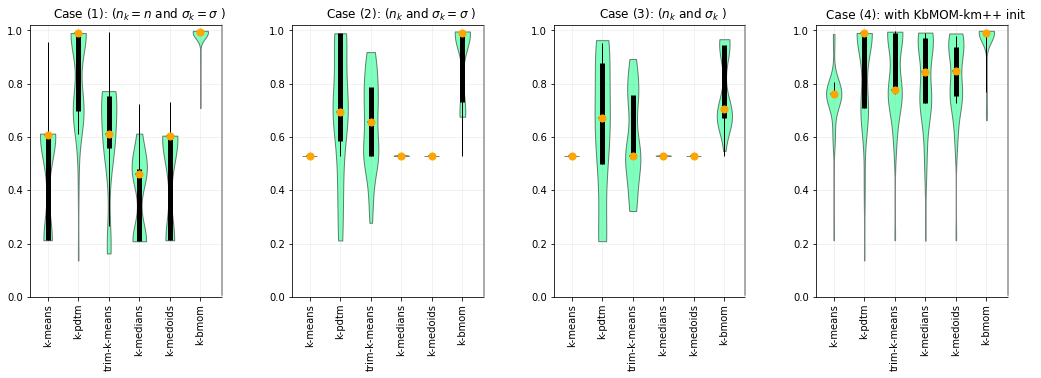}
\par\end{centering}
}
\par\end{centering}
\begin{centering}
\subfloat[{Violinplots of distortions focused on the window {[}500,12000{]}\label{fig:Benchmark_Violinplots_distortions} }]{\begin{centering}
\includegraphics[scale=0.41]{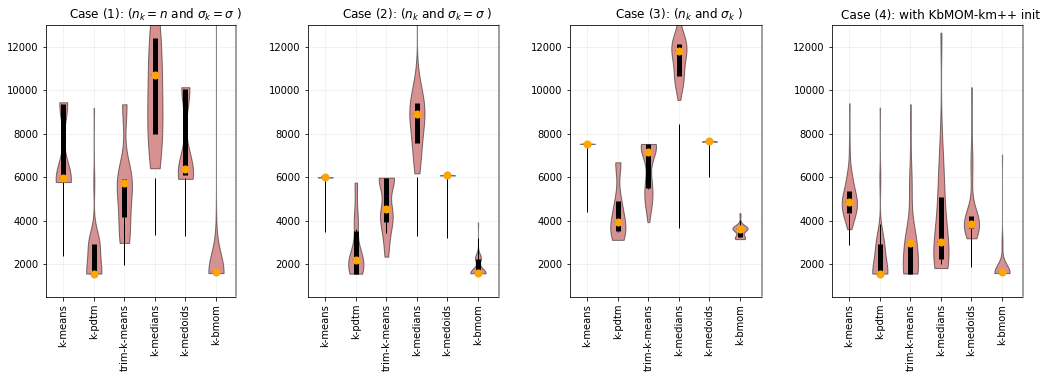}
\par\end{centering}
}
\par\end{centering}
\begin{centering}
\subfloat[Violinplots of the number of clusters found on the non polluted data
at the end of each procedure\label{fig:Benchmark_Violinplots_nbK}]{\begin{centering}
\includegraphics[scale=0.41]{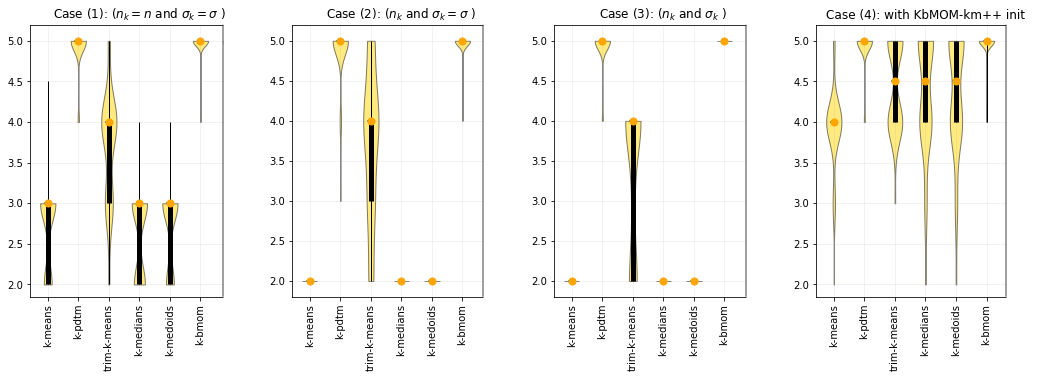}
\par\end{centering}
}
\par\end{centering}
\vspace{4ex}

\caption{Violinplots of different metrics computed on 50 repetitions of 5 kmeans-based
algorithms according to 3 frameworks. The median of each distribution
is depicted by an orange dot and the interquartile range by a thick
black vertical line.}
\end{figure}

The results of three simulated contexts presented above are summarized
in Tables~\ref{tab:Benchmark-Case-1}, \ref{tab:Benchmark_Case-2}
and \ref{tab:Benchmark_Case-3} where averages and standard deviations
of distortion, ARI and number of clusters describing the non polluted
data are displayed. Besides, the whole distribution of 50 repetitions
for each metric and tested algorithm are illustrated according to
violinplots in Figures~\ref{fig:Benchmark_Violinplots_ARI}, \ref{fig:Benchmark_Violinplots_distortions}
and \ref{fig:Benchmark_Violinplots_nbK} where the median of each
distribution is depicted by an orange dot and the interquartile range
by a thick black vertical line.

\begin{figure}
\begin{centering}
\subfloat[K-medians\label{fig:output_k-medians}]{\includegraphics[scale=0.39]{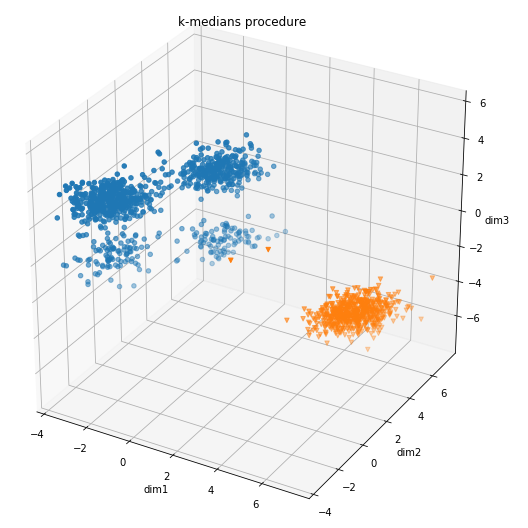}

}\hspace{3ex}\subfloat[trimmed K-means\label{fig:output_trimmed-k-means}]{\includegraphics[scale=0.39]{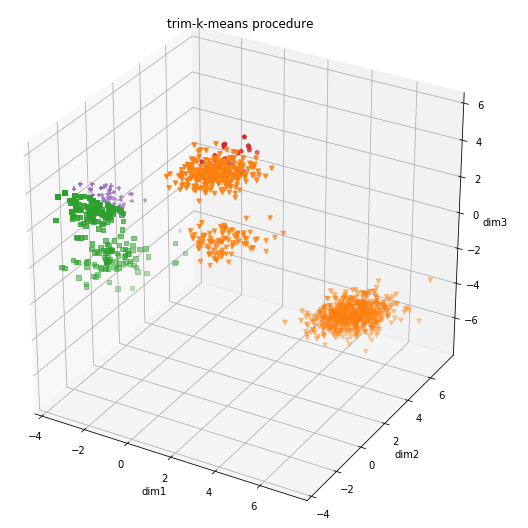}

}
\par\end{centering}
\begin{centering}
\subfloat[K-pdtm\label{fig:output_k-pdtm}]{\includegraphics[scale=0.4]{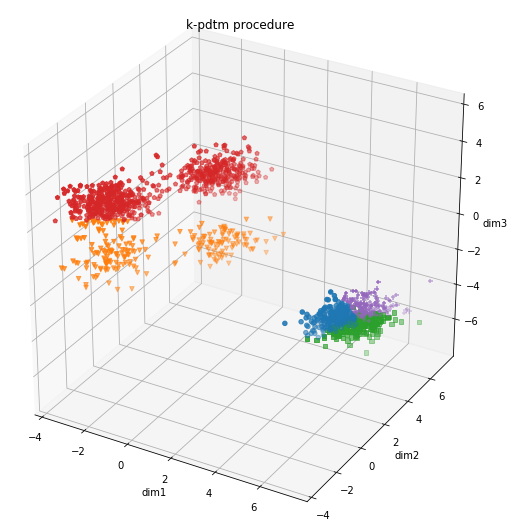}

}\hspace{3ex}\subfloat[K-bMOM\label{fig:output_k-mom}]{\includegraphics[scale=0.4]{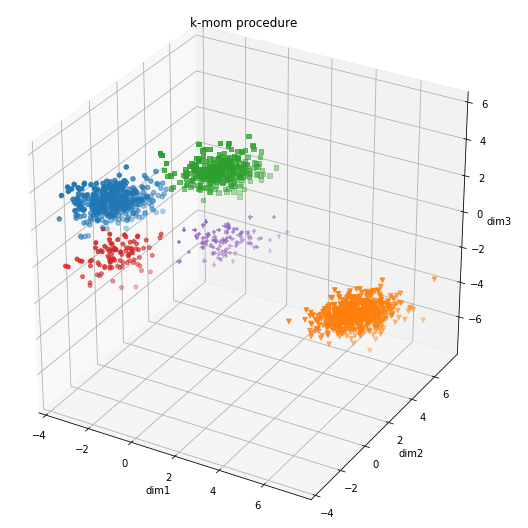}

}
\par\end{centering}
\caption{Examples of cluster assigment according to several procedures in the
more complex simulation case (unequal cluster size and unequal scaling
parameter in the covariance matrix). \textit{Note:} Outliers have
been removed from the pictures to ease the interpretation.}
\end{figure}

First of all, one can observe that the K-means, K-median and K-medoids
methods fail to discover the right number of clusters among the non
polluted data. Indeed, in average the outliers are grouped in 3 clusters
and the rest of the data in 2 instead of 5 groups in the first case
as it is illustrated in Table~\ref{tab:Benchmark-Case-1} and on
the violinplot in left side Figure~\ref{fig:Benchmark_Violinplots_nbK}.
Such a situation is partly explained by the initialisation process.
Indeed, the K-means++ procedure instantiates most of the time the
algorithm on one or two outliers. Thus, the Lloyd type algorithm whatever
is the agregation method used, is stucked in a local minima. This
situation gets worse in cases 2 and 3 since all 3 centers among 5
are located towards outliers as one can see in Table~\ref{tab:Benchmark_Case-2}
and Table~\ref{tab:Benchmark_Case-3} but also on the middle and
right side of Figure~\ref{fig:Benchmark_Violinplots_nbK} where the
associated violinplots can be summarized by a point. The cluster assignment
in the last context for the kmedians procedure is depicted in Figure~\ref{fig:output_k-medians}.
However, when the initial centers are well chosen, these 3 procedures
work better since half of the time, the true structure is discovered
for the robust versions as it shown in Table~\ref{tab:Benchmark-Case-1_2}.
The estimated centers become closest to their theoritical counterpart
especially for K-medians and K-medoids.

By looking at the number of clusters found among the non polluted
data, trimmed K-means and K-pdtm algorithms seem to have a better
behavior. K-pdtm tends to find the intrinsic structure all the times
(5 clusters) whatever is the situation considered since the average
number of clusters found among the non polluted data is around 4.9
in average with a very low standard deviation. Trimmed K-means most
of the time tends to put a center among the outliers since the number
among the non polluted data is quite stable and remains around 3.5.
However, the relevance of the data grouping decreases with the complexity
of the simulated situation and is really dependant of the algorithm.
Indeed, for the 3 simulated contexts, trimmed-K-means dispose of an
average ARI about 0.60 and an average distortion which is quite large
and reaches approximately 6000 \textit{ie} twice more than K-pdtm
distortion and almost equal to K-means distortion as we can observe
in Tables~\ref{tab:Benchmark-Case-1}, \ref{tab:Benchmark_Case-2}
and \ref{tab:Benchmark_Case-3}. The cluster assigment in Figure~\ref{fig:output_trimmed-k-means}
illustrates the failure of the algorithm to discover the true partition
of data.

On the other side, the ARI for K-pdtm reaches in average 0.88 in Table~\ref{tab:Benchmark-Case-1}.
Moreover, on the associated violinplot in the left side of Figure~\ref{fig:Benchmark_Violinplots_ARI},
we can see that this method is really performant since 50\% of the
time (the median is represented by an orange dot), the ARI on the
non polluted data is perfect and equals to 1 and the empirical distortion
is low. However, the performance of this method decreases and becomes
more erratic as the complexity of the situation increases. As we can
observe in the middle plots in Figure~\ref{fig:Benchmark_Violinplots_ARI},
the median ARI is as the same level as the average one which is about
0.71 and the distribution of ARI are spread almost uniformely between
0.5 and 1. An example of cluster assigment resulting from the K-pdtm
procedure after 300 iterations is depicted in Figure~\ref{fig:output_k-pdtm}.

Besides, even if the average performance tends to slowly decrease
according to the different situations, the proposed robust version
based on the MOM principle perform well in the presence of outliers.
Indeed, the intrinsic structure is almost all the time found in the
easiest context (Case 1) as the ARI, the distortions and the number
of clusters show it. The average ARI is up to 0.98 for the K-bMOM
algorithm with a standard deviation around 0.05 and the ARI median
reaches 1 as it is illustrated in the left hand side of Figure~\ref{fig:Benchmark_Violinplots_ARI}.
Moreover, the distortion is better and more stable for the K-bMOM
algorithm than the other versions as it can be observed in Table~\ref{tab:Benchmark-Case-1}
and Figure~\ref{fig:Benchmark_Violinplots_distortions}. This remark
remains true for the more complex contexts where the distortion is
more favorable for the K-bMOM algorithm both in average, in medians
and in variation. In the more constraint context (case 3), the K-bMOM
algorithm outperforms the rest of approaches even if the ARI is lower
and less stable than in the easiest simulated context as it can be
seen in the right, respectively left hand side of Figure~\ref{fig:Benchmark_Violinplots_ARI}.

Finally, when a same robust initialization is given to the robust
K-means based algorithms, as expected, the performances of the K-means,
trimmed K-means, K-median and K-medoids increase a lot : the partition
is better in average (up to 0.82) and also the overall distortion
which remain under 4000. In median, it can be observed in Figure~\ref{fig:Benchmark_Violinplots_ARI}
and \ref{fig:Benchmark_Violinplots_distortions} that K-median presents
better performance than K-medoids, trimmed K-means or K-means even
if it is less stable. The rest of approaches, K-pdtm and K-bMOM works
as well as in Case 1.

To conclude, this work provides a benchmark of robust K-means-based
clustering algorithms. Although it is still necessary to test their
performances on other different settings, our simulations give a preliminary
overview of performances of using MOM principle in clustering context.

Though the algorithmic principle of K-bMOM is the simplest one one
can think of when merging the Lloyd's algorithm and the Median-Of-Means
design, it has good performances compared to already known robust
K-means based algorithm in the presence of outliers.

\begin{table}[p]
\begin{centering}
\subfloat[Case 1 : equal cluster size and same covariance matrix ($\forall k,n_{k}=n$
and $\sigma_{k}=\sigma$)\label{tab:Benchmark-Case-1}]{\begin{centering}
\begin{tabular}{|l|ll|rr|ll|}
\hline 
\textsf{\small{}methods} & \multicolumn{2}{c|}{\textsf{\textbf{\small{}ari (std)}}} & \multicolumn{2}{c|}{\textsf{\textbf{\small{}distortion (std)}}} & \multicolumn{2}{c|}{\textsf{\textbf{\small{}nb groups (std)}}}\tabularnewline
\hline 
\hline 
\textsf{\small{}k-means} & \textsf{\small{}0.467} & \textsf{\small{}(0.185)} & \textsf{\small{}7096.1} & \textsf{\small{}(1650.0)} & \textsf{\small{}2.56} & \textsf{\small{}(0.49)}\tabularnewline
\textsf{\small{}k-pdtm} & \textsf{\small{}0.879} & \textsf{\small{}(0.176)} & \textsf{\textbf{\small{}2436.6}} & \textsf{\textbf{\small{}(1366.9)}} & \textsf{\textbf{\small{}4.90}} & \textsf{\textbf{\small{}(0.24)}}\tabularnewline
\textsf{\small{}trim-km} & \textsf{\small{}0.601} & \textsf{\small{}(0.180)} & \textsf{\small{}5375.4} & \textsf{\small{}(1949.7)} & \textsf{\small{}3.70} & \textsf{\small{}(0.60)}\tabularnewline
\textsf{\small{}k-median} & \textsf{\small{}0.378} & \textsf{\small{}(0.151)} & \textsf{\small{}11226.9} & \textsf{\small{}(3790.6)} & \textsf{\small{}2.52} & \textsf{\small{}(0.50)}\tabularnewline
\textsf{\small{}k-medoids} & \textsf{\small{}0.456} & \textsf{\small{}(0.178)} & \textsf{\small{}7536.8} & \textsf{\small{}(1846.2)} & \textsf{\small{}2.56} & \textsf{\small{}(0.49)}\tabularnewline
\hline 
\textsf{\small{}k-bmom} & \textsf{\textbf{\small{}0.981}} & \textsf{\textbf{\small{}(0.051)}} & \textsf{\textbf{\small{}1969.3}} & \textsf{\textbf{\small{}(1889.8)}} & \textsf{\textbf{\small{}4.98}} & \textsf{\textbf{\small{}(0.14)}}\tabularnewline
\hline 
\end{tabular}
\par\end{centering}

}
\par\end{centering}
\vspace{4ex}
\begin{centering}
\subfloat[Case 2 : unequal cluster size but same covariance matrix ($\forall k$,
$\sigma_{k}=\sigma$) \label{tab:Benchmark_Case-2}]{\begin{centering}
\begin{tabular}{|l|lr|rr|ll|}
\hline 
\textsf{\small{}methods} & \multicolumn{2}{c|}{\textsf{\textbf{\small{}ari (std)}}} & \multicolumn{2}{c|}{\textsf{\textbf{\small{}distortion (std)}}} & \multicolumn{2}{c|}{\textsf{\textbf{\small{}nb groups (std)}}}\tabularnewline
\hline 
\hline 
\textsf{\small{}k-means} & \textsf{\small{}0.529} & \textsf{\small{}(6.6e-16)} & \textsf{\small{}5998.4} & \textsf{\small{}(13.6)} & \textsf{\small{}2.00} & \textsf{\small{}(0.00)}\tabularnewline
\textsf{\small{}k-pdtm} & \textsf{\small{}0.704} & \textsf{\small{}(0.246)} & \textsf{\small{}2690.1} & \textsf{\small{}(1286.6)} & \textsf{\textbf{\small{}4.88}} & \textsf{\textbf{\small{}(0.32)}}\tabularnewline
\textsf{\small{}trim-km} & \textsf{\small{}0.656} & \textsf{\small{}(0.168)} & \textsf{\small{}4735.4} & \textsf{\small{}(1180.0)} & \textsf{\small{}3.74} & \textsf{\small{}(0.94)}\tabularnewline
\textsf{\small{}k-median} & \textsf{\small{}0.530} & \textsf{\small{}(0.002)} & \textsf{\small{}8746.6} & \textsf{\small{}(1777.7)} & \textsf{\small{}2.00} & \textsf{\small{}(0.00)}\tabularnewline
\textsf{\small{}k-medoids} & \textsf{\small{}0.529} & \textsf{\small{}(6.6e-16)} & \textsf{\small{}6100.2} & \textsf{\small{}(33.9)} & \textsf{\small{}2.00} & \textsf{\small{}(0.00)}\tabularnewline
\hline 
\textsf{\small{}k-bmom} & \textsf{\textbf{\small{}0.905}} & \textsf{\textbf{\small{}(0.131)}} & \textsf{\textbf{\small{}1843.3}} & \textsf{\textbf{\small{}(425.1)}} & \textsf{\textbf{\small{}4.98}} & \textsf{\textbf{\small{}(0.14)}}\tabularnewline
\hline 
\end{tabular}
\par\end{centering}
}
\par\end{centering}
\vspace{4ex}
\begin{centering}
\subfloat[Case 3 : unequal cluster size and different spherical covariance matrix
among clusters.\label{tab:Benchmark_Case-3}]{\begin{centering}
\begin{tabular}{|l|lr|rr|rl|}
\hline 
\textsf{\small{}methods} & \multicolumn{2}{c|}{\textsf{\textbf{\small{}ari (std)}}} & \multicolumn{2}{c|}{\textsf{\textbf{\small{}distortion (std)}}} & \multicolumn{2}{c|}{\textsf{\textbf{\small{}nb groups (std)}}}\tabularnewline
\hline 
\hline 
\textsf{\small{}k-means} & \textsf{\small{}0.529} & \textsf{\small{}(0.185)} & \textsf{\small{}7541.2} & \textsf{\small{}(9.8)} & \textsf{\small{}2.0} & \textsf{\small{}(0.0)}\tabularnewline
\textsf{\small{}k-pdtm} & \textsf{\small{}0.637} & \textsf{\small{}(0.176)} & \textsf{\small{}4397.7} & \textsf{\small{}(1206.3)} & \textsf{\small{}4.94} & \textsf{\small{}(0.23)}\tabularnewline
\textsf{\small{}trim-km} & \textsf{\small{}0.597} & \textsf{\small{}(0.110)} & \textsf{\small{}6460.0} & \textsf{\small{}(1141.2)} & \textsf{\small{}3.3} & \textsf{\small{}(0.56)}\tabularnewline
\textsf{\small{}k-median} & \textsf{\small{}0.530} & \textsf{\small{}(0.151)} & \textsf{\small{}11649.4} & \textsf{\small{}(1140.6)} & \textsf{\small{}2.0} & \textsf{\small{}(0.0)}\tabularnewline
\textsf{\small{}k-medoids} & \textsf{\small{}0.529} & \textsf{\small{}(0.178)} & \textsf{\small{}7651.5} & \textsf{\small{}(13.4)} & \textsf{\small{}2.0} & \textsf{\small{}(0.0)}\tabularnewline
\hline 
\textsf{\small{}k-bmom} & \textsf{\textbf{\small{}0.786}} & \textsf{\textbf{\small{}(0.134)}} & \textsf{\textbf{\small{}3516.5}} & \textsf{\textbf{\small{}(271.6)}} & \textsf{\textbf{\small{}5.0}} & \textsf{\textbf{\small{}(0.0)}}\tabularnewline
\hline 
\end{tabular}
\par\end{centering}
}
\par\end{centering}
\vspace{4ex}
\begin{centering}
\subfloat[Case 1 \textbf{with initialisation via KbMOM-km++ }(equal cluster
size and same covariance matrix)\ref{tab:Benchmark-Case-1_2}\label{tab:Benchmark-Case-1_2}]{\begin{centering}
\begin{tabular}{|l|ll|lr|rl|}
\hline 
\textsf{\small{}methods} & \multicolumn{2}{c|}{\textsf{\textbf{\small{}ari (std)}}} & \multicolumn{2}{c|}{\textsf{\textbf{\small{}distortion (std)}}} & \multicolumn{2}{c|}{\textsf{\textbf{\small{}nb groups (std)}}}\tabularnewline
\hline 
\hline 
\textsf{\small{}k-means} & \textsf{\small{}0.825} & \textsf{\small{}(0.185)} & \textsf{\small{}5069.3} & \textsf{\small{}(989.7)} & \textsf{\small{}3.92} & \textsf{\small{}(0.52)}\tabularnewline
\textsf{\small{}k-pdtm} & \textsf{\small{}0.877} & \textsf{\small{}(0.176)} & \textsf{\small{}2329.9} & \textsf{\small{}(1393.0)} & \textsf{\small{}4.96} & \textsf{\small{}(0.24)}\tabularnewline
\textsf{\small{}trim-km} & \textsf{\small{}0.820} & \textsf{\small{}(0.110)} & \textsf{\small{}2905.3} & \textsf{\small{}(1747.3)} & \textsf{\small{}4.44} & \textsf{\small{}(0.54)}\tabularnewline
\textsf{\small{}k-median} & \textsf{\small{}0.840} & \textsf{\small{}(0.151)} & \textsf{\small{}4001.7} & \textsf{\small{}(2453.1)} & \textsf{\small{}4.38} & \textsf{\small{}(0.72)}\tabularnewline
\textsf{\small{}k-medoids} & \textsf{\small{}0.841} & \textsf{\small{}(0.178)} & \textsf{\small{}4338.5} & \textsf{\small{}(1304.7)} & \textsf{\small{}4.38} & \textsf{\small{}(0.69)}\tabularnewline
\hline 
\textsf{\small{}k-bmom} & \textsf{\textbf{\small{}0.986}} & \textsf{\textbf{\small{}(0.074)}} & \textsf{\textbf{\small{}1808.1}} & \textsf{\textbf{\small{}( 814.4)}} & \textsf{\textbf{\small{}4.98}} & \textsf{\textbf{\small{}(0.14)}}\tabularnewline
\hline 
\end{tabular}
\par\end{centering}
}
\par\end{centering}
\vspace{4ex}

\caption{Distortions, ARI and number of clusters represented in the dataset
without outliers averaged among 50 repetitions of the K-means-based
approaches and their standard deviation according to 3 frameworks.\label{tab:Benchmark_Distortions_ARI}}
\end{table}
\pagebreak{}

\section{Color quantization in image processing\label{sec:Color-quantization}}

In this last experimental section, the K-bMOM procedure is applied
to the problem of color quantization adressed in image processing
and computer graphics. 

Color quantization (CQ) is a process which aims at reducing the number
of colors used in an image with the goal to keep the same quality
of visualisation as the original one. CQ is a challenging problem
since most of real-world images contain tens of thousands of colors.
Moreover such a procedure is commonly used ; it is indeed applied
for different tasks such as color analysis, compression, segmentation,
non-photorealistic rendering, and so one (see REF for ). 

CQ can be viewed as a clustering problem in 3-dimensions according
to the Red, Green, Blue channels of pixels of an image. A wide literature
is devoted to this problem and it appears that the K-means algorithm
is not used so often because of its sensitivity to the initialization
and computations requirements. We propose therefore to use the K-bMOM
procedure as a robust CQ process providing confident and high-quality
quantization on a bunch of popular images. As benchmark, the obtained
results are compared to the comparative study driven by~\cite{thompson2019fast}
on 17 CQ procedures well-known in the literature.

\subsection{Images and experimental setup}

\begin{figure}[H]
\begin{centering}
\subfloat[Baboon]{\includegraphics[scale=0.2]{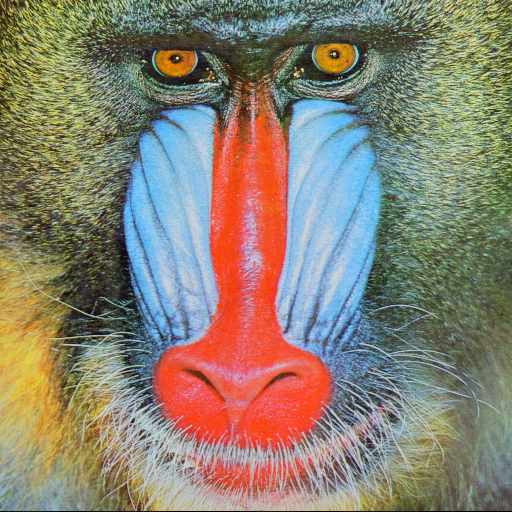}

}\subfloat[Peppers]{\includegraphics[scale=0.2]{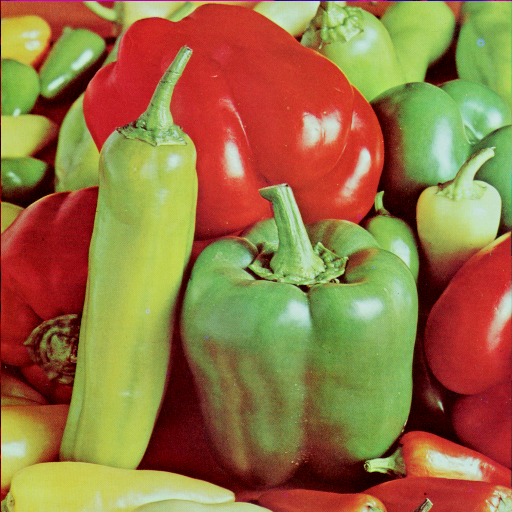}

}\subfloat[Parrots]{\includegraphics[scale=0.2]{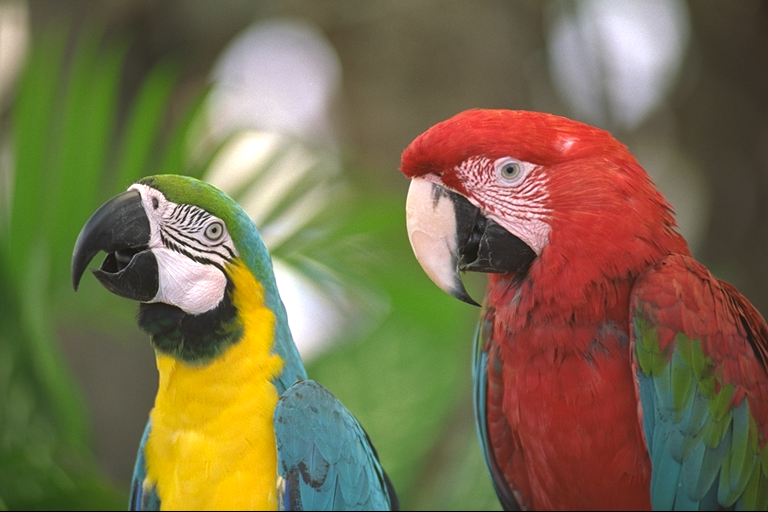}

}
\par\end{centering}
\caption{Images used to process Color Quantization with K-bMOM\label{fig:Images-application}}

\end{figure}

The K-bMOM method has been tested on 3 popular 24-bit test images
-- Baboon (512 $\times$ 512), Peppers (512 $\times$ 512), and Parrots
(768 $\times$ 512)-- which are detailed in Table~\ref{tab:Details-of-studied-images}
and illustrated in Figure~\ref{fig:Images-application}:
\begin{center}
\begin{table}[H]
\begin{centering}
\begin{tabular}{l|c|c|l}
\hline 
\multicolumn{1}{l}{\textsf{\small{}name}} & \multicolumn{1}{c}{\textsf{\small{}size}} & \multicolumn{1}{c}{\textsf{\small{}unique colors}} & \textsf{\small{}source}\tabularnewline
\hline 
\textsf{\small{}Baboon} & \textsf{\small{}$512\times512\times3$} & \textsf{\small{}$230\,427$} & \textsf{\small{}USC-SIPI Image Database}\tabularnewline
\hline 
\textsf{\small{}Peppers} & \textsf{\small{}$512\times512\times3$} & \textsf{\small{}$183\,525$} & \textsf{\small{}USC-SIPI Image Database}\tabularnewline
\hline 
\textsf{\small{}Parrots} & \textsf{\small{}$768\times512\times3$} & \textsf{\small{}$72\,079$} & \textsf{\small{}Kodak Lossless True Color Image Suite}\tabularnewline
\hline 
\end{tabular}
\par\end{centering}
\caption{Details of studied images from the USC-SIPI and Kodak Lossless True
Color Image Suite Databases.\label{tab:Details-of-studied-images}}
\end{table}
\par\end{center}

Each image has been reshapen into a matrix of $w\times512$ pixels
($w\in\{512,768\}$) with 3-dimensions linked to RGB channels. The
K-bMOM algorithm has been repeated 50 times on each image for a number
of colors (or clusters) equals to 32, 64 and 128 respectively. For
these 3 segmentations, the number of blocks have been set to $B=200$,
the size of each block set to $n_{B}=2000$ and the number maximum
of iterations have been fixed to 50. 

\subsection{Experimental results}

In order to evaluate the quality of the quantization, the Mean Square
Error have been computed between the pixels of the original image
and their segmented version, then averaged among 50 repetitions. The
standard deviation is also computed. Besides, in order to compare
the results obtained by K-bMOM with the well-known versions in the
image processing litterature, we display the minimum MSE obtained
in the recent literature on these images and the $95$th percentile
(for more details on results and proposed CQ methods, see~\cite{thompson2019fast}).
The results are summarized in Table~\ref{tab:MSE_CQ}.

\begin{table}[H]
\begin{centering}
\begin{tabular}{|l|cc|cc|cc|}
\cline{2-7} 
\multicolumn{1}{l|}{} & \multicolumn{2}{c|}{$K=32$} & \multicolumn{2}{c|}{$K=64$} & \multicolumn{2}{c|}{$K=128$}\tabularnewline
\multicolumn{1}{l|}{\textsf{\small{}Image }} & \textsf{K-bMOM} & \textsf{benchmark} & \textsf{K-bMOM} & \textsf{benchmark} & \textsf{K-bMOM} & \textsf{benchmark}\tabularnewline
\hline 
\textsf{\small{}Baboon} & $377$ \textit{(1.3)} & $[374,643]$ & $238$ \textit{(0.6)} & $[234,445]$ & $155$ \textit{(0.5)} & $[149,307]$\tabularnewline
\textsf{\small{}Peppers} & $231$ \textit{(1.8)} & $[249,418]$ & $135$ \textit{(2.1)} & $[148,257]$ & $86$ \textit{(2.2)} & $[88,174]$\tabularnewline
\textsf{\small{}Parrots} & $234$ \textit{(5.5)} & $[231,441]$ & $126$ \textit{(0.9)} & $[127,265]$ & $77$ \textit{(0.6)} & $[73,154]$\tabularnewline
\hline 
\end{tabular}
\par\end{centering}
\caption{Average and standard deviation (in parenthesis) of MSE obtained by
the K-bMOM procedure for $K=\{32,64,128\}$ colors. In brackets, the
minimum and the percentile 95 of MSE obtained on a benchmark of CQ
methods in~\cite{thompson2019fast}.\label{tab:MSE_CQ}}
\end{table}
First of all, it can be noted that color quantization processed by
the K-bMOM approach competes with the best CQ methods in terms of
quality. Indeed, the average MSE are on the same order of magnitude
as the minimum MSE obtained on benchmark CQ algorithms. In some cases,
as for instance on Peppers image with $K\in\{32,64\}$, K-bMOM procedure
presents the lowest MSE i.e. the best quality for image color representation.
Moreover, the procedure remains very stable which guarantees the efficiency
of the procedure compared to a traditional K-means algorithm.

\begin{figure}
\begin{centering}
\subfloat[Baboon output images]{\begin{centering}
\includegraphics[scale=0.45]{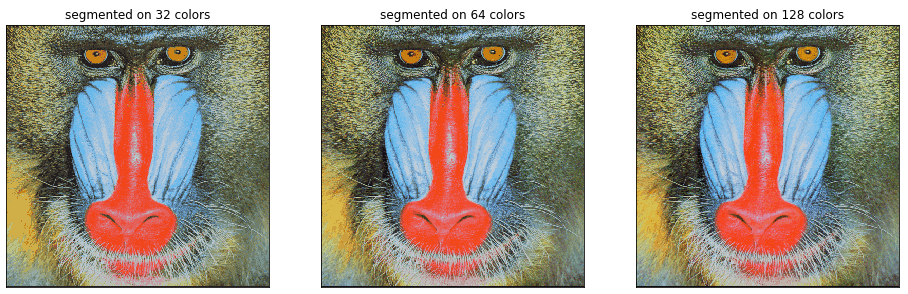}
\par\end{centering}
}
\par\end{centering}
\begin{centering}
\subfloat[Peppers output images]{\begin{centering}
\includegraphics[scale=0.45]{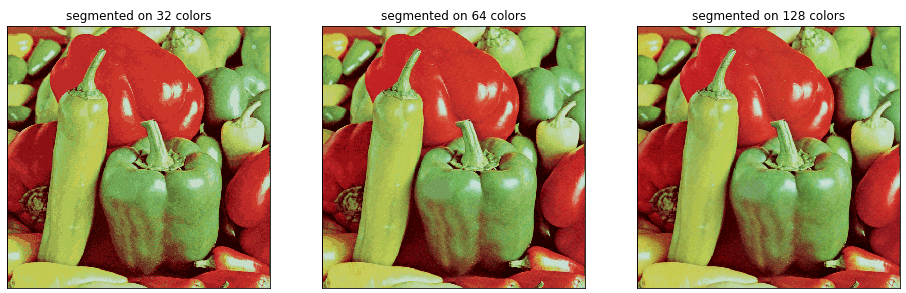}
\par\end{centering}
}
\par\end{centering}
\begin{centering}
\subfloat[Parrots output images]{\begin{centering}
\includegraphics[scale=0.45]{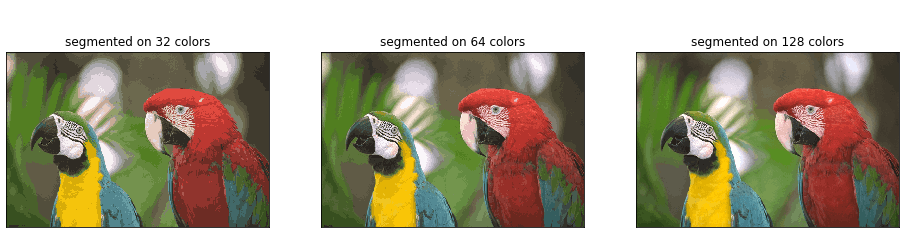}
\par\end{centering}

}
\par\end{centering}
\caption{Sample quantization results for $K=32$, $64$ and $128$ respectively
from left to right on Baboon, Peppers and Parrots images. \label{fig:color_quantized_images}}
\end{figure}

\begin{figure}
\begin{centering}
\subfloat[Baboon error images.]{\begin{centering}
\includegraphics[scale=0.45]{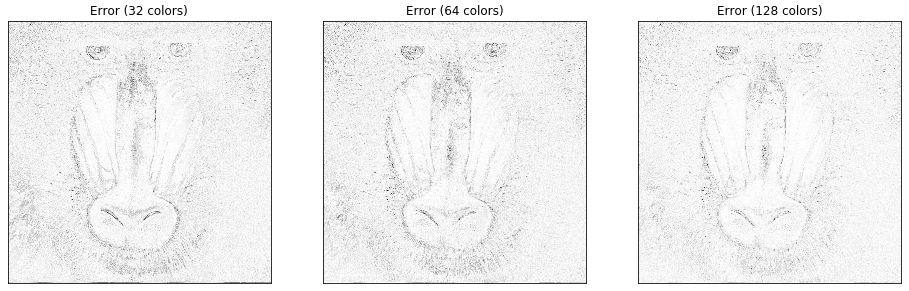}
\par\end{centering}

}
\par\end{centering}
\begin{centering}
\subfloat[Peppers error images.]{\begin{centering}
\includegraphics[scale=0.45]{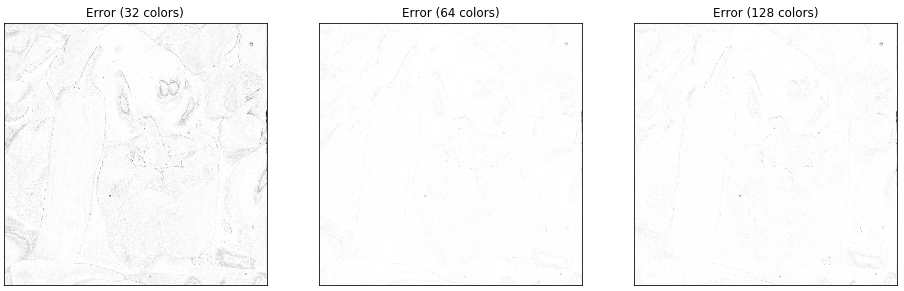}
\par\end{centering}

}
\par\end{centering}
\begin{centering}
\subfloat[Parrots error images.]{\begin{centering}
\includegraphics[scale=0.45]{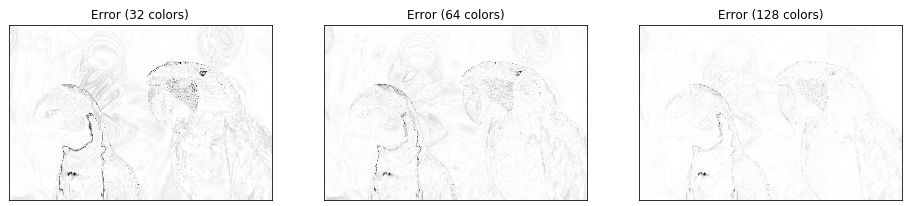}
\par\end{centering}

}
\par\end{centering}
\caption{Full scale error images for $K=32$, $64$ and $128$ respectively
from left to right on Baboon, Peppers and Parrots images. \label{fig:Error-images}}
\end{figure}

Besides, Figures~\ref{fig:color_quantized_images}a., \ref{fig:color_quantized_images}b.
and \ref{fig:color_quantized_images}c. illustrate the quantization
process on Baboons, Peppers and Parrots images for $K=32,$ $64$
and $128$ respectively. Figures~\ref{fig:Error-images}a., \ref{fig:Error-images}b.
and \ref{fig:Error-images}c. show the squared error per pixel in
a reverse gray scaled mapping which means that the higher is the MSE,
the darker is the pixel. It can be seen that the K-bMOM approach performs
very well in allocating $K-$representative colors to the different
image regions, especially on peppers where the error images are really
clean.


\section{Proof of Theorem \ref{thm:perf_bound}\label{sec:Proof-of-Theorem}}

Assume without loss of generality that $B\geq8$ (otherwise the bound
stated in Theorem \ref{thm:perf_bound} may occur with probability
zero). We have, by definition of $\hat{\mathbf{c}}_{n}$, for any
constant $a>0$,
\begin{align*}
 & \mathbb{P}\left(R\left(\hat{\mathbf{c}}_{n}\right)-R_{*}>a\right)\\
\leq & \mathbb{P}\left(\inf_{\mathbf{c}\in\mathcal{F}_{>a}}{\rm MOM}\left(\ell_{\mathbf{c}}\right)\leq\inf_{\mathbf{c}\in\mathcal{F}_{a}}{\rm MOM}\left(\ell_{\mathbf{c}}\right)\right)\\
= & \mathbb{P}\left(\sup_{\mathbf{c}\in\mathcal{F}_{>a}}\left\{ R_{*}-{\rm MOM}\left(\ell_{\mathbf{c}}\right)\right\} \geq\sup_{\mathbf{c}\in\mathcal{F}_{a}}\left\{ R_{*}-{\rm MOM}\left(\ell_{\mathbf{c}}\right)\right\} \right)\\
\leq & \mathbb{P}\left(\sup_{\mathbf{c}\in\mathcal{F}_{>a}}\left\{ R_{*}-{\rm MOM}\left(\ell_{\mathbf{c}}\right)\right\} \geq R_{*}-{\rm MOM}\left(\ell_{\mathbf{c}_{*}}\right)\right),
\end{align*}
where $\mathcal{F}_{a}=\left\{ \mathbf{c}\in\mathcal{X}_{M_{*}}^{k}:R(\mathbf{c})-R_{*}\leq a\right\} $
and $\mathcal{F}_{>a}=\left\{ \mathbf{c}\in\mathcal{X}_{M_{*}}^{k}:R(\mathbf{c})-R_{*}>a\right\} =\mathcal{X}_{M_{*}}^{k}\left\backslash \mathcal{F}_{a}\right..$
Now, on the one hand, for any $x>0$,
\begin{align*}
 & \mathbb{P}\left({\rm MOM}\left(\ell_{\mathbf{c}_{*}}\right)-R_{*}\geq x\right)\\
= & \mathbb{P}\left(\sum_{j=1}^{B}\mathbf{1}_{\left\{ \left(P_{b_{j}}-P\right)\left(\ell_{\mathbf{c}_{*}}\right)\geq x\right\} }\geq\frac{B}{2}\right)\\
\leq & \mathbb{P}\left(\sum_{j\in I}\mathbf{1}_{\left\{ \left(P_{b_{j}}-P\right)\left(\ell_{\mathbf{c}_{*}}\right)\geq x\right\} }\geq\frac{B}{2}-|O|\right)\\
= & \sum_{j=\left\lfloor B/2-|O|\right\rfloor }^{B}\left(\begin{array}{c}
B\\
j
\end{array}\right)p^{j}(1-p)^{B-j}\\
 & \leq p^{\left\lfloor B/2-|O|\right\rfloor }2^{B}
\end{align*}
where $p=\mathbb{P}\left(\left(P_{b_{j}}-P\right)\left(\ell_{\mathbf{c}_{*}}\right)\geq x\right)$.
In addition, by Markov inequality,
\[
p\leq\frac{B{\rm Var\left(\ell_{\mathbf{c}_{*}}\right)}}{nx^{2}}.
\]
Hence, by choosing $x=\sqrt{64eB{\rm Var}\left(\ell_{\mathbf{c}_{*}}\right)/n}$,
we get 
\[
\mathbb{P}\left({\rm MOM}\left(\ell_{\mathbf{c}_{*}}\right)-R_{*}\geq x\right)\leq2^{B}\left(\frac{1}{64e}\right)^{\left\lfloor B/2-|O|\right\rfloor }.
\]
Note that since $|O|\leq B/4$ and $B\geq8$, we have $\left\lfloor B/2-|O|\right\rfloor \geq\left\lfloor B/4\right\rfloor \geq B/8$
and $2^{B}\leq16^{\left\lfloor B/4\right\rfloor +1}\leq64^{\left\lfloor B/4\right\rfloor }$.
This gives
\[
\mathbb{P}\left({\rm MOM}\left(\ell_{\mathbf{c}_{*}}\right)-R_{*}\geq x\right)\leq\exp\left(-\frac{B}{8}\right).
\]
On the other hand, 
\begin{align*}
 & \mathbb{P}\left(\sup_{\mathbf{c}\in\mathcal{F}_{>a}}\left\{ R_{*}-{\rm MOM}\left(\ell_{\mathbf{c}}\right)\right\} \geq-x\right)\\
\\
\leq & \mathbb{P}\left(\sup_{\mathbf{c}\in\mathcal{F}_{>a}}\left\{ \frac{1}{B}\sum_{j=1}^{B}\mathbf{1}_{\left\{ R_{*}-P_{b_{j}}\left(\ell_{\mathbf{c}}\right)\geq-x\right\} }\right\} \geq\frac{1}{2}\right)\\
\leq & \mathbb{P}\left(\sup_{\mathbf{c}\in\mathcal{F}_{>a}}\left\{ \frac{1}{|I|}\sum_{j\in I}\mathbf{1}_{\left\{ R_{*}-P_{b_{j}}\left(\ell_{\mathbf{c}}\right)\geq-x\right\} }\right\} \geq\frac{B}{2|I|}-\frac{|O|}{|I|}\right)\\
\end{align*}

Let us denote $\Delta=B/2|I|-|O|/|I|$. Let us now recall the self-bounding
condition (see \cite[Theorem 6.12]{BouLugMas:13}).
\begin{defn}
A function $f$ is said to have the \textit{self-bounding property}
if, for some functions $f_{i}:\mathcal{Z}^{n-1}\rightarrow\mathbb{R}$,
for all $z=\left(z_{1},...,z_{n}\right)\in\mathcal{Z}^{n}$ and for
all $i=1,...,n$,
\[
0\leq f\left(z\right)-f_{i}\left(z^{\left(i\right)}\right)\leq1
\]
and
\[
\sum_{i=1}^{n}\left(f\left(z\right)-f_{i}\left(z^{\left(i\right)}\right)\right)\leq f\left(z\right)\text{ ,}
\]
where $z^{\left(i\right)}=\left(z_{1},...,z_{i-1},z_{i+1},....,z_{n}\right)$. 
\end{defn}
\begin{lem}
\label{prop_self_bounding}If $\mathcal{A}$ is a class of sets on
a measurable space $\left(\mathcal{Z},\mathcal{T}\right)$, then the
function $h:\mathcal{Z}^{p}\rightarrow\mathbb{R}$ defined by 
\[
h\left(z_{1},...,z_{p}\right)=\sup_{A\in\mathcal{A}}\sum_{j=1}^{p}1_{A}\left(z_{j}\right)\text{ ,}
\]
has the self-bounding property. By consequence, if $\left(\xi_{1},...,\xi_{p}\right)\in\mathcal{X}^{p}$
is an i.i.d. sample, then by setting $Z=h\left(\xi_{1},...,\xi_{p}\right)$,
it holds for any $t>0$,
\begin{equation}
\mathbb{P}\left(Z\geq\mathbb{E}Z+t\right)\leq\exp\left(-\frac{t^{2}}{2\mathbb{E}Z+2t/3}\right)\text{ .}\label{ineq_concen_self_bounding}
\end{equation}
\end{lem}
\bigskip{}

\begin{proof}
Denote $h_{i}\left(z^{\left(i\right)}\right)=\sup_{A\in\mathcal{A}}\sum_{j\neq i}1_{A}\left(z_{j}\right)$
. Then
\[
0\leq h\left(z\right)-h_{i}\left(z^{\left(i\right)}\right)\leq\sup_{A\in\mathcal{A}}1_{A}\left(z_{i}\right)\leq1\text{ .}
\]
Also, assume without loss of generality that $h\left(z\right)=\sum_{j=1}^{I}1_{A_{\ast}\left(z\right)}\left(z_{j}\right)$
for some $A_{\ast}\left(z\right)\in\mathcal{A}$, then 
\begin{eqnarray*}
\sum_{i=1}^{I}\left(h\left(z\right)-h_{i}\left(z^{\left(i\right)}\right)\right) & \leq & \sum_{i=1}^{I}\left(\sum_{j=1}^{I}1_{A_{\ast}\left(z\right)}\left(z_{j}\right)-\sum_{j\neq i}1_{A_{\ast}\left(z\right)}\left(z_{j}\right)\right)\\
 & = & \sum_{i=1}^{I}1_{A_{\ast}\left(z\right)}\left(z_{i}\right)=h\left(z\right)\text{ .}
\end{eqnarray*}
Hence, $h$ has the self-bounding property. Now, inequality (\ref{ineq_concen_self_bounding})
simply follows from \cite[Theorem 6.12]{BouLugMas:13}. 
\end{proof}
\begin{cor}
The following process
\[
Z\left(\mathcal{F}_{>a},x\right)=\sup_{\mathbf{c}\in\mathcal{F}_{>a}}\frac{1}{|I|}\sum_{j\in I}\mathbf{1}_{\left\{ R_{*}-P_{b_{j}}\left(\ell_{\mathbf{c}}\right)\geq-x\right\} }
\]
 is concentrated around its expected value according to the following
inequality, 
\begin{equation}
\mathbb{P}\left(Z\left(\mathcal{F}_{>a},x\right)\geq\Delta\right)\leq\exp\left(-\frac{|I|(\Delta-\mathbb{E}\left[Z\left(\mathcal{F}_{>a},x\right)\right])^{2}}{2\mathbb{E}\left[Z\left(\mathcal{F}_{>a},x\right)\right]+2(\Delta-\mathbb{E}\left[Z\left(\mathcal{F}_{>a},x\right)\right])/3}\right)\text{ .}\label{eq:concen_Z}
\end{equation}
\end{cor}
\begin{proof}
It suffices to apply Lemma \ref{prop_self_bounding} with $p=n_{B}$,
$\mathcal{Z}=\mathcal{X}^{n_{B}}$, $\xi_{i}=\left(X_{j}\right)_{j\in b_{i}}$
for $i\in I$ and
\[
\mathcal{A}=\left\{ \left\{ z=\left(x_{1},...,x_{n_{B}}\right):\frac{-1}{\left\vert n_{B}\right\vert }\sum_{j=1}^{n_{B}}\ell_{\mathbf{c}}\left(x_{j}\right)+R_{*}>-x\right\} :\mathbf{c}\in\mathcal{F}_{>a}\right\} \text{ .}
\]
\end{proof}
Consider a function $\phi:\mathbb{R}\rightarrow\mathbb{R}$, such
that $\phi(t)=(t-1)1_{\left\{ 1\leq t\leq2\right\} }+1_{\left\{ t\geq2\right\} }.$
The function $\phi$ is thus $1$-Lipschitz and it holds $\phi(t)\geq$
$1_{\left\{ t\geq2\right\} }$. Therefore, 
\begin{align}
\mathbb{E}\left[Z\left(\mathcal{F}_{>a},x\right)\right] & =\mathbb{E}\left[\sup_{\mathbf{c}\in\mathcal{F}_{>a}}\left\{ \frac{1}{|I|}\sum_{j\in I}\mathbf{1}_{\left\{ \left(P-P_{b_{j}}\right)\left(\ell_{\mathbf{c}}\right)\geq R\left(\mathbf{c}\right)-R_{*}-x\right\} }\right\} \right]\nonumber \\
\leq & \mathbb{E}\left[\sup_{\mathbf{c}\in\mathcal{F}_{>a}}\left\{ \frac{1}{|I|}\sum_{j\in I}\mathbf{1}_{\left\{ \left(P-P_{b_{j}}\right)\left(\ell_{\mathbf{c}}\right)\geq a-x\right\} }\right\} \right]\nonumber \\
\leq & \mathbb{E}\left[\sup_{\mathbf{c}\in\mathcal{F}_{>a}}\left\{ \frac{1}{|I|}\sum_{j\in I}\phi\left(\frac{2\left(P-P_{b_{j}}\right)\left(\ell_{\mathbf{c}}\right)}{a-x}\right)\right\} \right]\label{eq:Z}\\
\nonumber \\
\nonumber \\
\nonumber 
\end{align}
Now, for any $i\in I$,
\[
\mathbb{E}\left[\phi\left(\frac{2\left(P-P_{b_{i}}\right)\left(\ell_{\mathbf{c}}\right)}{a-x}\right)\right]\leq\mathbb{P}\left[\left(P-P_{b_{i}}\right)\left(\ell_{\mathbf{c}}\right)\geq(a-x)/2\right]\leq\frac{BL}{n\left(a-x\right)^{2}},
\]
where the constant $L$ is such that $\sup_{c}{\rm Var}\left(\ell_{c}\right)\leq L$.
More explicitly, we can choose $L=16M^{2}\mathbb{E}\left[\Vert X\Vert^{2}\right].$
Hence, by Inequality (\ref{eq:Z}) we get,
\[
\mathbb{E}\left[Z\left(\mathcal{F}_{>a},x\right)\right]\leq\frac{BL}{n\left(a-x\right)^{2}}+\mathbb{E}\left[\sup_{\mathbf{c}\in\mathcal{F}_{>a}}\left\{ \frac{1}{|I|}\sum_{j\in I}\phi\left(\frac{2\left(P-P_{b_{j}}\right)\left(\ell_{\mathbf{c}}\right)}{a-x}\right)-\mathbb{E}\left[\phi\left(\frac{2\left(P-P_{b_{j}}\right)\left(\ell_{\mathbf{c}}\right)}{a-x}\right)\right]\right\} \right].
\]
Now, by a standard symmetrisation argument, it holds
\begin{align*}
 & \mathbb{E}\left[\sup_{\mathbf{c}\in\mathcal{F}_{>a}}\left\{ \frac{1}{|I|}\sum_{j\in I}\phi\left(\frac{2\left(P-P_{b_{j}}\right)\left(\ell_{\mathbf{c}}\right)}{a-x}\right)-\mathbb{E}\left[\phi\left(\frac{2\left(P-P_{b_{j}}\right)\left(\ell_{\mathbf{c}}\right)}{a-x}\right)\right]\right\} \right]\\
\leq2 & \mathbb{E}\left[\sup_{\mathbf{c}\in\mathcal{F}_{>a}}\left\{ \frac{1}{|I|}\sum_{j\in I}\epsilon_{j}\phi\left(\frac{2\left(P-P_{b_{j}}\right)\left(\ell_{\mathbf{c}}\right)}{a-x}\right)\right\} \right],
\end{align*}
where the $\epsilon_{j}$'s are i.i.d. Rademacher variables (i.e.
$\mathbb{P}(\epsilon_{j}=1)=\mathbb{P}(\epsilon_{j}=-1)=1/2$) independent
from the sample. Furthermore, as the function $\phi$ is $1$-Lipschitz
and $\phi(0)=0$, we can apply the so-called contraction principle,
which gives
\begin{align*}
 & \mathbb{E}\left[\sup_{\mathbf{c}\in\mathcal{F}_{>a}}\left\{ \frac{1}{|I|}\sum_{j\in I}\epsilon_{j}\phi\left(\frac{2\left(P-P_{b_{j}}\right)\left(\ell_{\mathbf{c}}\right)}{a-x}\right)\right\} \right]\\
\leq & \frac{2}{a-x}\mathbb{E}\left[\sup_{\mathbf{c}\in\mathcal{F}_{>a}}\left\{ \frac{1}{|I|}\sum_{j\in I}\epsilon_{j}\left(P-P_{b_{j}}\right)\left(\ell_{\mathbf{c}}\right)\right\} \right]
\end{align*}
and by symmetrisation again,
\begin{align*}
 & \mathbb{E}\left[\sup_{\mathbf{c}\in\mathcal{F}_{>a}}\left\{ \frac{1}{|I|}\sum_{j\in I}\epsilon_{j}\left(P-P_{b_{j}}\right)\left(\ell_{\mathbf{c}}\right)\right\} \right]\\
\leq & \frac{2B}{|I|n}\mathbb{E}\left[\sup_{\mathbf{c}\in\mathcal{F}_{>a}}\left\{ \sum_{i\in\mathcal{J}}\epsilon_{i}\ell_{\mathbf{c}}\left(X_{i}\right)\right\} \right],
\end{align*}
where $\mathcal{J}=\bigcup_{j\in I}b_{j}$. By Lemma 4.3 in \cite{MR2444554},
\begin{align*}
\mathbb{E}\left[\sup_{\mathbf{c}\in\mathcal{F}_{>a}}\left\{ \sum_{i\in\mathcal{J}}\epsilon_{i}\ell_{\mathbf{c}}\left(X_{i}\right)\right\} \right] & \leq2k\sqrt{\vert\mathcal{J}\vert}\left[M\sqrt{\mathbb{E}\left[\Vert X\Vert^{2}\right]}+M^{2}/2\right].\\
 & \leq2k\sqrt{n}\left[M\sqrt{\mathbb{E}\left[\Vert X\Vert^{2}\right]}+M^{2}/2\right]
\end{align*}
Putting things together, we obtain
\[
\mathbb{E}\left[Z\left(\mathcal{F}_{>a},x\right)\right]\leq\frac{BL}{n\left(a-x\right)^{2}}+\frac{8B}{\left(a-x\right)|I|\sqrt{n}}2k\left[M\sqrt{\mathbb{E}\left[\Vert X\Vert^{2}\right]}+M^{2}/2\right].
\]
Now, by taking
\begin{equation}
a\geq\max\left\{ 2x,4\sqrt{\frac{BL}{n\Delta}},\frac{128Bk\left[M\sqrt{\mathbb{E}\left[\Vert X\Vert^{2}\right]}+M^{2}/2\right]}{\Delta|I|\sqrt{n}}\right\} ,\label{eq:def_a_1}
\end{equation}
we get 
\[
\frac{BL}{n\left(a-x\right)^{2}}\leq\frac{\Delta}{4}
\]
and 
\[
\frac{8B}{\left(a-x\right)|I|\sqrt{n}}2k\left[M\sqrt{\mathbb{E}\left[\Vert X\Vert^{2}\right]}+M^{2}/2\right]\leq\frac{\Delta}{4}.
\]
This gives $\mathbb{E}\left[Z\left(\mathcal{F}_{>a},x\right)\right]\leq\Delta/2$
and so, by using Inequality (\ref{eq:concen_Z}),
\[
\mathbb{P}\left(Z\left(\mathcal{F}_{>a},x\right)\geq\Delta\right)\leq\exp\left(-\frac{3|I|\Delta}{16}\right).
\]
To conclude, it suffices now to notice that if $n_{o}\leq B/4$, then
$|O|\leq B/4$, $|I|\geq3B/4$ and $\Delta\geq B/(4|I|)\geq1/4.$
Indeed, in this case, Inequality (\ref{eq:def_a_1}) is achieved by
choosing for instance
\[
a=\max\left\{ 8\sqrt{\frac{eBL}{n}},512\frac{k\left[M\sqrt{\mathbb{E}\left[\Vert X\Vert^{2}\right]}+M^{2}/2\right]}{\sqrt{n}}\right\} .
\]

\pagebreak{}

\bibliographystyle{plain}
\bibliography{bibliKMOM,Slope_heuristics_regression_13}

\end{document}